\documentstyle[11pt,aaspp4]{article}

\begin{document}

\slugcomment{\em To appear in ApJ, April 1, 1997, Vol. 478}

\title{Gravitational Lensing of the X-Ray Background \\ 
by Clusters of Galaxies}
\author{Alexandre Refregier\altaffilmark{1}}
\affil{Columbia Astrophysics Laboratory, 538 W. 120th Street,
New York, NY 10027 \\ email:refreg@odyssey.phys.columbia.edu}
\and
\author{Abraham Loeb}
\affil{Astronomy Department, Harvard University, Cambridge, MA 02138 \\
email:aloeb@cfa.harvard.edu}

\altaffiltext{1}{also at the Department of Physics, Columbia University}

\begin{abstract}
Gravitational lensing by clusters of galaxies affects the cosmic X-ray
background (XRB) by altering the observed density and flux distribution of
background X-ray sources.  At faint detection flux thresholds, the resolved
X-ray sources appear brighter and diluted, while the unresolved component
of the XRB appears dimmer and more anisotropic, due to lensing. The diffuse
X-ray intensity in the outer halos of clusters might be lower than the
sky-averaged XRB, after the subtraction of resolved sources. Detection of
the lensing signal with a wide-field X-ray telescope could probe the mass
distribution of a cluster out to its virialization boundary.  In
particular, we show that the lensing signature imprinted on the resolved
component of the XRB by the cluster A1689, should be difficult but possible
to detect out to $8^\prime$ at the $2$--$4\sigma$ level, after $10^6$
seconds of observation with the forthcoming AXAF satellite. The lensing
signal is fairly insensitive to the lens redshift in the range $0.1\la
z_l\la0.6$. The amplitude of the lensing signal is however sensitive to the
faint end slope of the number-flux relation for unresolved X-ray sources,
and can thus help constrain models of the XRB.  A search for X-ray arcs or
arclets could identify the fraction of all faint sources which originate
from extended emission of distant galaxies.  The probability for a
3$\sigma$ detection of an arclet which is stretched by a factor $\sim 3$
after a $10^6$ seconds observation of A1689 with AXAF, is roughly
comparable to the fraction of all background X-ray sources that have an
intrinsic size $\sim 1^{\prime\prime}$.
\end{abstract}


\section{Introduction}
The current epoch in the evolution of structure in the universe is
marked by the formation of clusters of galaxies. Since clusters trace the
transition between collapse and virialization, their internal structure and
evolution offer a test bed for a critical examination of popular
cosmological theories (e.g., Richstone, Loeb, \& Turner 1992; Bahcall \& Cen
1994; Cen \& Ostriker 1994; Eke, Cole, \& Frenk 1996).

The structure of clusters has been traditionally studied by observing
their X-ray emission and the kinematics of their galaxies
(e.g. Sarazin 1988; Forman \& Jones 1990).  More recently, deep
optical imaging has allowed the use of gravitational lensing of faint
galaxies to independently map the cluster mass distribution. The
lensing study can be performed either in the cluster core--where the
lensing signature is strong, or well outside the core--where lensing
is weak (see reviews by Schneider, Ehlers, \& Falco 1992; Schneider
1995; Kaiser 1995; and Narayan \& Bartelmann 1996). Aside from
introducing distortions to the images of extended sources, lensing
changes the statistics of both the resolved (Broadhurst, Taylor, \&
Peacock 1995) and the unresolved (Waerbeke et al. 1996; Waerbeke \&
Mellier 1996) components of the extragalactic optical background.
While the traditional X-ray and dynamical methods infer the cluster mass
distribution by assuming spherical symmetry and virial equilibrium,
the lensing techniques can measure the projected surface density
distribution of the cluster free of these assumptions.  The lensing
methods can thus uncover deviations from spherical symmetry and
equilibrium, especially in the outer regions of clusters where
detection of the cluster X-ray emission is difficult.

In the current paper, we examine the imprint of lensing by galaxy
clusters on the resolved and unresolved components of the cosmic X-ray
background (XRB) (see also Refregier \& Loeb 1996a for a brief
discussion). The effect of lensing on the resolved component of the
XRB typically dominates over the cluster emission at a radius of a few
Mpc from the cluster center. Thus, X-ray telescopes with large fields
of view ($\sim 10^\prime$) can probe the mass distribution of a
cluster, possibly out to its virialization boundary. The information
provided by the lensing method about the cluster envelope is
complementary to the information obtained from the traditional X-ray
emission method about the inner region of the cluster.  A single deep
X-ray observation can be used to study the cluster potential through
both methods simultaneously.

The magnification due to lensing brings into view X-ray sources that
are otherwise too faint to be resolved.  With the aid of galaxy
clusters as {\it natural} telescopes, one can thus reach greater
sensitivity to the faint number-flux relation of extragalactic X-ray
sources, and by that, unravel the origin of the X-ray background.  In
this work, we demonstrate, through the specific example of the cluster
A1689, that lensing by individual clusters could possibly be studied with
the forthcoming {\it Advanced X-ray Astronomy Facility} (AXAF)
satellite, which is scheduled for launch in 1998 (Weisskopf et al. 1987;
Elvis et al. 1995).

The paper is organized as follows. In \S\ref{clusters} we approximate
cluster lenses as singular isothermal spheres and model their X-ray
emission phenomenologically. The cluster emission provides photon
noise that masks the lensing signal in the cluster core, but declines
rapidly at larger radii. We consider in detail the specific example of
A1689 which we use as a test case for our study. In \S\ref{XRB}, we
model the XRB by considering various extrapolations of the faint end
of the number-flux relation for extragalactic X-ray sources beyond
that directly observed with the ROSAT satellite. In
\S\ref{effect_lensing}, we examine the effect of lensing on these
sources and on the unresolved component of the XRB. In
\S\ref{observability}, we discuss the detectability of the lensing
signature for both the resolved and the unresolved components of the
XRB.  Finally, we summarize our conclusions in \S\ref{conclusion}.
Unless specified otherwise, we assume a cosmological density parameter
$\Omega=1$ and a Hubble constant $H_{0}=50$ km s$^{-1}$ Mpc$^{-1}$.
\label{cosmo_params}

\section{Clusters}
\label{clusters}
\subsection{Lensing Model}
\label{model_cluster_potential}
In our lensing calculations, we model a cluster of galaxies as a
Singular Isothermal Sphere (SIS) (see e.g., \markcite{sch92}Schneider
et al. 1992).  This model provides a good first-order approximation to
the projected mass distribution of known cluster lenses (Tyson \&
Fischer 1995; \markcite{nar96}Narayan \& Bartelmann 1996;
\markcite{sq96a}\markcite{sq96b}Squires et al. 1996a,b).

The surface mass density of a SIS is given by
\begin{equation}
\Sigma(\xi) = \frac{\sigma_{v}^{2}}{2G\xi}
\label{eq:sigma_sis}
\end{equation}
where $G$ is Newton's constant, $\sigma_{v}$ is the line-of-sight
velocity dispersion of the lens, and $\xi$ is the projected
radius. This distribution has a diverging mass as $\xi \rightarrow
\infty$ and should therefore be truncated at a finite value of $\xi$,
of order the virialization boundary of the cluster ($\sim 5 {\rm
Mpc}$).  The singularity of $\Sigma(\xi)$ at $\xi=0$ can be removed
through the addition of a core (see e.g.
\markcite{mir95}Miralda-Escud\'{e} \& Babul 1995;
\markcite{dai96}Daines et al. 1996). As long as the core radius is
smaller than about half of the Einstein radius (which characterizes
the scale of the strong lensing zone, see below), the lensing
properties of the system do not differ significantly from those of the
associated SIS. This is the case in many of the known cluster lenses
(Kneib \& Soucail 1996; Narayan \& Bartelmann 1996), including A1689
which will be used as a test case in our analysis (see \S\ref{a1689}).

Next, let us consider a point source behind the SIS and denote the
angle between the unlensed source location on the sky and the cluster
center by $\hat{\theta}$. The lensing effect of the cluster causes the
image of the source to be displaced, magnified, and sometimes
split. For a SIS, the angle of an image relative to the cluster
center, $\theta$, is obtained from the lens equation
(\markcite{sch92}Schneider et al. 1992),
\begin{equation}
\hat{\theta} = \theta - \frac{\theta}{|\theta|} \alpha,
\label{eq:lensing}
\end{equation}
where negative angles refer to positions on the opposite side of the
cluster center. The Einstein angle $\alpha$ is given by
\begin{equation}
\alpha(z_l,z_s) \equiv 4 \pi \left( \frac{\sigma_{v}}{c} \right)^{2}
  \frac{D_{ls}}{D_{os}},
\label{eq:einstein_angle}
\end{equation}
where $c$ is the speed of light, and $D_{os}$ and $D_{ls}$ are the
angular-diameter distances between the observer and the source,
and the lens and the source, respectively. These distances
depend on the source and lens redshifts, and on the cosmological
parameters $\Omega$ and $H_{0}$  (see, e.g. \markcite{ko92}Kochanek 1992).

\begin{figure}
\plotone{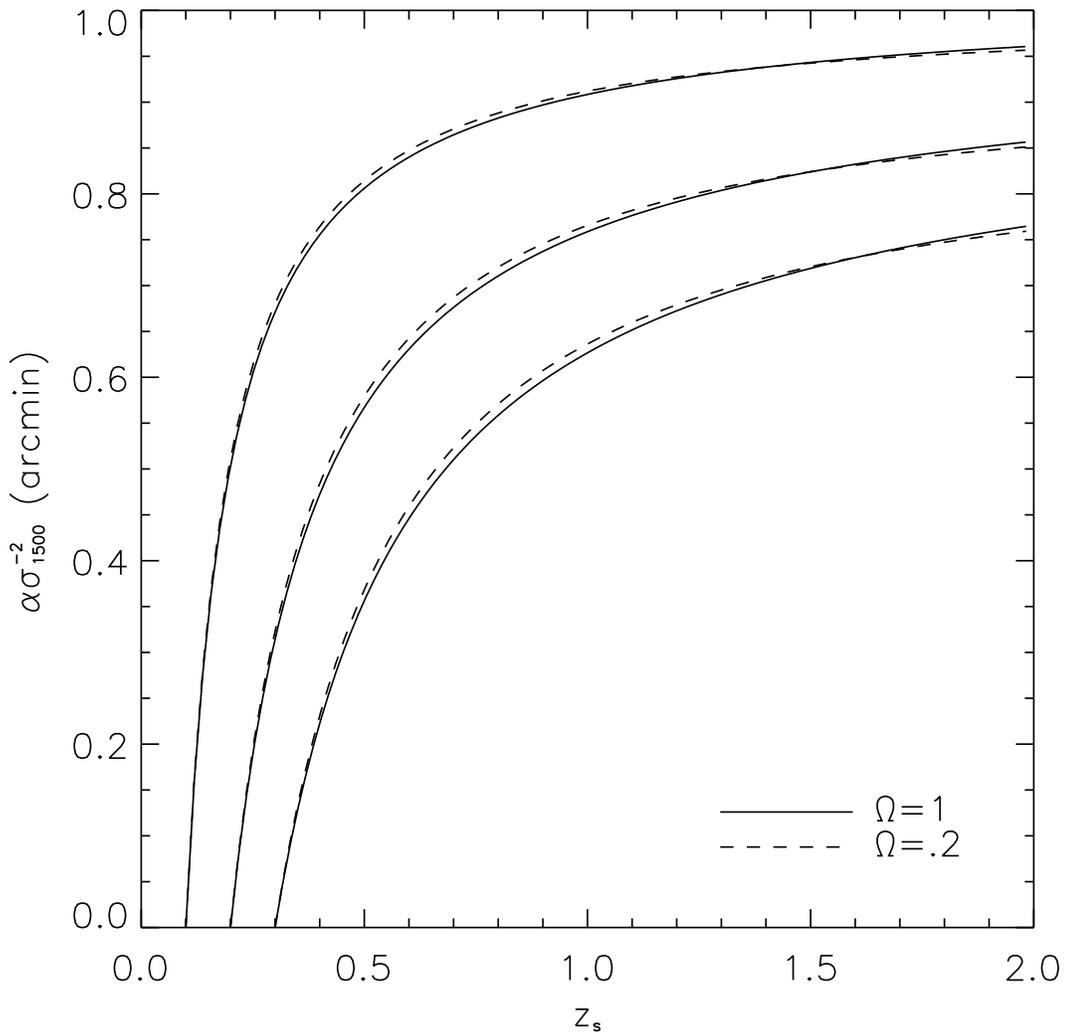}
\caption{Einstein angle $\alpha$ as a function of source redshift
$z_{s}$ for a SIS.  The Einstein angle is scaled by $\sigma_{1500}
\equiv \sigma_{v}/(1500$ km s$^{-1})$, where $\sigma_{v}$ is the
1D velocity dispersion.  The solid and dashed lines correspond to a
cosmological density parameter of $\Omega=1$ and 0.2, respectively.
The lines from left to right correspond to lens redshifts of
$z_{l}=0.1,0.2$ and 0.3, respectively.
\label{fig:alpha_zs}}
\end{figure}

Equation~(\ref{eq:lensing}) implies that when $|\hat{\theta}| <
\alpha$, the source has two images on opposite sides of the cluster
center. For $|\hat{\theta}|>\alpha$, only one image is present.
Figure~\ref{fig:alpha_zs} shows the value of the Einstein angle
$\alpha$ as a function of $z_{s}$ for $z_{l}=0.1,0.2,0.3$, and for
$\Omega=0.2,1$ with a zero cosmological constant. The Einstein angle
is larger for nearby clusters but depends only weakly on $\Omega$. As
noted in \S\ref{cosmo_params}, we will assume $\Omega=1$ throughout
the rest of the paper.

\placefigure{fig:alpha_zs}

A source with an unlensed flux $\hat{S}$ acquires a flux
$S(\theta)=\mu(\theta) \hat{S}$ due to lensing. The magnification
$\mu$ depends on the image position $\theta$ as
\begin{equation}
\mu(\theta)= \left[ 1- \left| \frac{\alpha}{\theta} \right| \right]^{-1}.
\label{eq:mu_theta}
\end{equation}
Negative values of $\mu$ correspond to inverted images. For large
values of $\theta$, $\mu \approx 1$ and the source is weakly affected
by the lensing potential. On the other hand, for $\theta=\alpha$ (or
$\hat{\theta}=0$), the magnification diverges.  In practice, the
maximum magnification is limited by the finite extent of the lensed
source. A small source which is perfectly aligned with the cluster
center produces an ``Einstein ring'' at $\theta=\alpha$. More often,
imperfect alignments or deviations of the cluster potential from
axial symmetry around the line-of-sight, result in partial rings or
``arcs'' near the same radius.  Note that for an image at
$\theta<\frac{\alpha}{2}$, $\vert\mu\vert$ is smaller than unity, and
so the image is de-magnified relative to the unlensed source.

\subsection{X-Ray Emission from the Lensing Cluster}
The X-ray emission of the lensing cluster limits the ability of an
X-ray telescope to resolve faint sources behind it. It must therefore
be taken into account when the detectability of the lensing signal is
examined.  Most of the X-ray luminosity of clusters originates from
bremsstrahlung emission by the hot intra-cluster gas (for reviews, see
Forman \& Jones 1982; \markcite{sar88}Sarazin 1988).  In principle,
one could self-consistently derive the X-ray intensity profile
expected from the SIS potential, assuming that the gas is isothermal
and in hydrostatic equilibrium.

However, detailed X-ray studies often reveal a more complicated
situation.  In some clusters, such as A2218 or A1689, preliminary
lensing studies do not reproduce the cluster mass estimate based on
X-ray observations under the assumption of hydrostatic equilibrium
(Miralda-Escud\'e \& Babul 1995; Loeb \& Mao 1994).  Although the
inner core of clusters is expected to relax to equilibrium on a short
dynamical time scale ($\la 10^9 ~{\rm yr}$), the outer parts of clusters
could be disturbed for longer periods as a result of mergers or
anisotropic infall of gas.  Recent ROSAT and ASCA observations show
temperature gradients in rich X-ray clusters (Briel \& Henry 1994;
\markcite{hen95}Henry \& Briel 1995; \markcite{he96}Henriksen \&
Markevitch 1996; \markcite{ma96}Markevitch 1996), and sometimes require the
use of a multiphase analysis regarding the X-ray emitting gas
(\markcite{all96}Allen et al. 1996).  Rich clusters with deep potential
wells which provide strong gravitational lensing signatures, often contain
cooling flows where the simple single-phase analysis of the X-ray data is
inadequate.  Aside from kinematic departures from hydrostatic equilibrium,
the low value of the gas density in the outer parts of clusters allows for
the possibility that the electron temperature measured from the X-ray
emission, be lower than the ion temperature. This effect is expected to be
particularly important in the rarefied regions near the virialization shock
of the infalling gas, where the ions--which carry most of the
inertia--thermalize their bulk motion and become hotter than the electrons
(Fox \& Loeb 1996).  A potential sign of this effect was, in fact, implied
from recent ASCA observations of A2163 (Markevitch et al.  1996).

A theoretical analysis of the above issues is beyond the scope of this
paper. For the purposes of this study, we instead adopt a
phenomenological approach and consider models of the X-ray emission of
clusters based on direct observations.

\subsection{Test Cases}
\subsubsection{A1689}
\label{a1689}
For concreteness, we first consider the case of the cluster A1689,
which is located at a redshift $z_{l} = 0.181$. In the optical band,
faint elongated arcs have been observed at $\theta_{\rm arc} \approx
51^{\prime\prime}$ from the cluster center (Tyson \& Fisher 1995). The
Einstein radius of A1689, $\alpha \approx \theta_{\rm arc}$, is one of
the largest observed in clusters (Le F\`{e}vre et al. 1994; Kneib \&
Soucail 1996). In addition, the projected appearance of the cluster is
regular, aside from a secondary substructure which is located $\sim1'$
($\approx240$ kpc) to the north-east of the center, which we ignore
here. These two facts make A1689 a good candidate for our study.

The inversion of the weak-lensing distortions of optical galaxies
behind A1689 results in a mass profile which is close to an isothermal
sphere (i.e., with $\Sigma \propto \theta^{-1}$) for $\theta \lesssim
3'.5$ (Tyson \& Fischer 1995).  Although there is some marginal
evidence that the profile steepens at $\theta \gtrsim 3'.5$, we take
the potential to be isothermal at all radii and model the cluster as a
SIS (Eq.~[\ref{eq:sigma_sis}]). Under the assumption that
$\alpha(z_l,z_s \rightarrow \infty) \approx \theta_{arc}$
(cf. Eq.~[\ref{eq:einstein_angle}]), the observed arc radius implies a
line-of-sight velocity dispersion $\sigma_{v} \approx 1390$ km
s$^{-1}$.

Miralda-Escud\'{e} \& Babul (1995) performed a detailed modeling of
A1689 assuming two isothermal mass clumps with cores. The best fit
parameters for the primary clump include a velocity dispersion of
$\sigma_v = 1450 $ km s$^{-1}$ and a core radius of
$\theta_{core}=10^{\prime\prime}$.  (The much smaller secondary mass
clump has $\sigma_v = 700$ km s$^{-1}$ and
$\theta_{core}=10^{\prime\prime}$ and is located close to the position
of the substructure which we neglected above.) The presence of a core
reduces the mass inside the Einstein radius and thus requires a
velocity dispersion which is slightly higher than our SIS
estimate. Note that the core is well within the Einstein radius and
can therefore be neglected for our purposes. The direct measurement of
the velocity dispersion of the galaxies in the direction of the
cluster yields an unreasonably high value, $\sigma_v \approx 2355$ km
s$^{-1}$ (Teague et al.  1990), probably due to projection of
secondary clumps at different redshifts along the line-of-sight (Tyson
\& Fischer 1995; Daines et al. 1995).

The X-ray properties of A1689 have been studied with ROSAT by Daines
et al.  (1996), who combined their result with earlier GINGA
observations by Arnaud et al. (1994).  The temperature structure is
complicated by the presence of a cooling flow, which was resolved by
ROSAT. Here, we only consider the single temperature thermal plasma
model for the spectrum of the cluster X-ray emission.  The best fit
temperature in the $1$--$5$ arcmin annulus of the ROSAT observation is
$kT \approx 10.5$ keV with a metalicity of 35\% the solar value and
with an HI column density $N_{H} \approx 1.5 \times 10^{20}$
cm$^{-2}$. The 2--10 keV flux is about $1.72 \times 10^{-11}$ ergs
cm$^{-2}$ s$^{-1}$, corresponding to an intrinsic source luminosity of
$L_{x}=2.85 \times 10^{45}$ ergs s$^{-1}$.

In the outer parts of the cluster, the intensity distribution is well
described by a King profile (King 1962), but for $\theta \lesssim
50^{\prime\prime}$ the intensity is significantly above the prediction
of the King model.  An acceptable fit to the full 0.4--2 keV profile is
provided by the functional form (Daines et al. 1996),
\begin{equation}
i_{c}(0.4-2\mbox{keV},\theta) \approx i_{c,0} 
\left( \frac{\theta}{\theta_{x}} \right)^{-\beta_x}
\left( 1+\frac{\theta}{\theta_{x}} \right)^{\beta_{x}-\beta_{x}'},
\label{eq:i_daines}
\end{equation}
where $i_{c,0} \approx 1.8 \times 10^{-12}$ ergs
cm$^{-2}$ s$^{-1}$ arcmin$^{-2}$, $\theta_{x} \approx 2'.33$,
$\beta_{x} \approx 0.53$, and $\beta_{x}' \approx 5.19$. 

\subsubsection{Displaced A1689}
\label{a1689_displaced}
As another example, we consider a cluster like A1689 but displaced to
a different redshift $z_{l}$. This example is useful for studying the
dependence of the lensing signal on the cluster redshift. While the
velocity dispersion $\sigma_{v}$ of the cluster is left unchanged, the
Einstein angle $\alpha$ varies according to
equation~(\ref{eq:einstein_angle}). 

The observed X-ray brightness profile of the cluster also depends on
the cluster redshift. The observed specific intensity,
$({di}/{d\epsilon})$, is related to that at the source,
$({di^{\prime}}/{d\epsilon})$, by
$\frac{di}{d\epsilon}(\epsilon,\theta)= (1+z)^{-3}
\times 
\frac{di^{\prime}}{d\epsilon}[\epsilon(1+z),\xi(\theta,z)]$, where
$\epsilon$ is the observed photon energy, $\theta$ is the angle from the
cluster center, and $z$ is the cluster redshift.  The quantity $\xi=\theta
D_{ol}(z)$ is the projected radius at the source corresponding to an
observed angle $\theta$, where $D_{ol}$ is the angular-diameter distance
between the observer and the cluster.  We approximate the energy dependence
of the bremsstrahlung intensity of the cluster as,
$\frac{di^{\prime}}{d\epsilon}(\epsilon,\xi) \propto e^{-\epsilon/kT}$,
where $T$ is the cluster temperature and $k$ is Boltzmann's constant. It
then follows that the observed intensity $i_c$ of the displaced cluster in
the energy range $(\epsilon_{1},\epsilon_{2})$ is related to the actual
intensity $i_a$ of A1689 observed in the same band, through the relation
\begin{equation}
i_{c}(\theta) \approx 
i_{a}\!\!\left[\theta \frac{D_{ol}(z_{l})}{D_{ol}(z_{a})} \right]
\left( \frac{1+z_a}{1+z_l} \right)^4
\frac{e^{-\epsilon_1/kT_l}-e^{-\epsilon_2/kT_l}}
     {e^{-\epsilon_1/kT_a}-e^{-\epsilon_2/kT_a}},
\end{equation}
where $z_a\equiv0.181$ is the actual redshift of A1689, and
$kT_{j}\equiv kT/(1+z_j)$ for $j=l,a$. In the 0.4-2keV range, $i_a$ is
given by equation~(\ref{eq:i_daines}).

\section{X-Ray Background}
\label{XRB}
\subsection{Observational Facts}
Since its discovery by Giacconi et al. (1962), the XRB has been the
subject of numerous observational studies and considerable theoretical
debates (for reviews, see Fabian \& Barcons 1992; Zamorani 1995; De
Zotti et al. 1995; and Hasinger 1996a,b). The COBE limits on the
distortion of the spectrum of the Cosmic Microwave Background
precludes the possibility that the XRB originates from a homogeneous
hot intergalactic plasma (Mather et al. 1990; Fixsen et al.  1996),
despite the similarity between its spectrum and a thermal
bremsstrahlung spectrum.  The alternative explanation, namely a
superposition of discrete sources, is supported by deep ROSAT
observations, which identified $\approx60\%$ of the soft (1-2 keV) XRB
as being due to resolved point sources (Hasinger et al. 1993;
Vikhlinin et al. 1995a,b,c). Hasinger et al. (1993) place an upper
limit of about 25\% for a truly diffuse component of the XRB in this
band. Even though a sufficiently clumped intergalactic plasma, of the
type observed in large-scale numerical simulations of structure
formation (Cen et al. 1995), could still contribute to the XRB without
conflicting with the COBE limits (Loeb \& Ostriker 1992), its
contribution is probably small, especially above 2 keV. Note that
while discrete sources can acquire strong magnifications, any diffuse
emission, even if moderately clumped, is only weakly affected by lensing.

Optical identifications of sources in the ROSAT survey (see Hasinger
1996b for a recent summary) imply that about 60\% of X-ray sources
brighter than $10^{-14}$ ergs cm$^{-2}$ s$^{-1}$ in the 0.5--2 keV
band are AGN. The nature of the fainter source population is still
unknown.  Recent deep surveys (Jones et al. 1995; Boyle et al. 1995;
Carballo et al.  1995; Griffiths et al. 1996; but see also Hasinger
1996a,b; Refregier et al.  1996)
suggest that narrow-emission-line galaxies can become important at
faint fluxes. Since the X-ray emission in AGN originates from their
compact cores, these sources can be regarded as point-like.
Consequently, gravitational lensing could magnify them considerably,
but is not likely to produce detectable arcs or arclets.  On the other
hand, the finite extent of galaxies would result in more limited
observable magnifications but could lead to the appearance of X-ray
arcs and arclets.  The search for X-ray arcs can therefore be used to
test the nature of the faint population of extragalactic X-ray
sources.
\label{galaxies}

The spectrum of the XRB was most recently observed by ASCA (Gendreau
et al. 1995) and was found to have the form,
\begin{equation}
\frac{di_{\rm XRB}}{d\epsilon} \approx 9.6 \epsilon^{-0.41} 
\mbox{keVcm$^{-2}$ s$^{-1}$ sr$^{-1}$ keV$^{-1}$}
\label{eq:xrb_spectrum}
\end{equation}
in the 1-7 keV range, where $\epsilon$ is the photon energy in keV.
This spectrum is somewhat inconsistent with earlier measurements with
other instruments, including ROSAT (for a discussion of this
discrepancy see Hasinger 1996a; Chen et al. 1995). For our purposes,
we will use the spectrum of equation~(\ref{eq:xrb_spectrum}) except
when modeling the source counts (see \S\ref{counts_model} below),
where the ROSAT measurement is more appropriate.  The above XRB
spectrum is harder than expected for ROSAT sources with $S(0.5-2
\mbox{keV}) \gtrsim 10^{-14}$ ergs cm$^{-2}$ s$^{-1}$ whose spectral
index is $\sim1$ (the ``spectral paradox'').  Below this flux, the
source spectral index smoothly flattens to a value $\sim0.4$
(Vikhlinin et al. 1995c). This value is indeed required in order to
match the ROSAT to the ASCA source counts (Inoue et
al. 1996).  We therefore adopt a mean spectral index for the
background X-ray sources of $\gamma_b \approx 0.4$.
\label{xrb_spectrum}

\subsection{Models for the X-Ray Background}
To model the XRB, we only consider its discrete component. For now, we
neglect the finite angular size of the sources and model the XRB as a
collection of point-sources. In \S\ref{arcs}, we discuss the effect of
the possible finite size of sources, which could arise if galaxies
were found to contribute a significant fraction of the XRB.
As a first approximation, we neglect the clustering of X-ray
sources (see Vikhlinin \& Forman 1995), and thus take the sources to
be randomly distributed in space.
\label{neglect_clustering}

The Einstein angle $\alpha$ in equation~(\ref{eq:einstein_angle})
depends on the source redshift $z_{s}$.  However,
figure~\ref{fig:alpha_zs} implies that $\alpha$ is only a weak
function of $z_{s}$ for $z_s \gtrsim 3 z_{l}$. Thus, for a cluster
with $z_{s} \sim 0.2$ such as A1689, $\alpha$ is almost independent of
$z_s$ for $z_s \gtrsim 0.6$. The deep ROSAT survey by Boyle et
al. (1993) revealed that AGN with $S(.5-2\mbox{keV})>6 \times
10^{-15}$ ergs cm$^{-2}$ s$^{-1}$ have a mean redshift of $z_{s} \sim
1.5$. We therefore ignore, to a first approximation, the redshift
distribution of the background X-ray sources and assume $z_s
\rightarrow \infty$ for the purpose of calculating $\alpha$.
If faint sources turned out to be dominated by
galaxies, and if their mean redshift were considerably smaller than
that of AGN, then one would need to take the source redshift
distribution into account. In this discussion we also neglect the
enhancement in the source counts due to discrete sources associated
with the cluster.  Because of the proximity of the cluster, these
sources (most likely embedded AGN, such as the one found in Cl0016+16
[Neumann \& B\"ohringer 1996]) are likely to have optical counterparts
or obvious galactic hosts with measurable redshifts, and can therefore
be separated from the lensed source population.

Following the above simplifications, we only need to specify a model
for the number-flux distribution of faint X-ray sources. For this
purpose, we extend the deepest observed counts from ROSAT beyond the
ROSAT detection threshold. We model the differential counts ${dn/dS}$
as three broken power laws,
\begin{equation}
\left. \frac{dn}{dS} \right|_{S} = \left\{ \begin{array}{ll}
                          \eta_1 S^{-\beta_1}, & S>S_{12} \\
                          \eta_2 S^{-\beta_2}, & S_{12}>S>S_{23} \\
                          \eta_3 S^{-\beta_3}, & S<S_{23},
                           \end{array}
                   \right.
\label{eq:model-scalings}
\end{equation}
where $n$ is the number of sources per square degree and S is the X-ray
flux in the ROSAT band (0.5-2 keV) in ergs cm$^{-2}$ s$^{-1}$.  We then
impose the following observational constraints based on the ROSAT
survey of Hasinger et al. (1993):
\begin{enumerate}
\item{} The counts must agree with the ROSAT counts for fluxes in 
  the 0.5-2 keV band which are
  brighter than $S_{\rm ROSAT}= 2.66 \times 10^{-14}$ ergs
  cm$^{-2}$ s$^{-1}$.
\item{} The counts must be within the fluctuation analysis
  limits derived from considering the anisotropies of the
  residual XRB observed by ROSAT. 
\item{} The total integrated intensity of the sources
  must be less than the XRB intensity,
  $i_{\rm XRB}(0.5-2 \mbox{keV}) \approx 7.61 \times 10^{-12}$ ergs cm$^{-2}$
  s$^{-1}$ deg$^{-2}$.
\end{enumerate}
The value for $i_{\rm XRB}(0.5-2 \mbox{keV})$ was extrapolated from the
estimate of the extragalactic component of the XRB in the 1-2 keV band
by Hasinger et al. (1993), assuming a power law spectrum with an index
of 1. This is more appropriate than the somewhat lower intensity
resulting from an integration of equation~(\ref{eq:xrb_spectrum}),
which was derived with a different instrument (see discussion in
\S\ref{xrb_spectrum}).
\label{counts_model}

\begin{figure}
\plotone{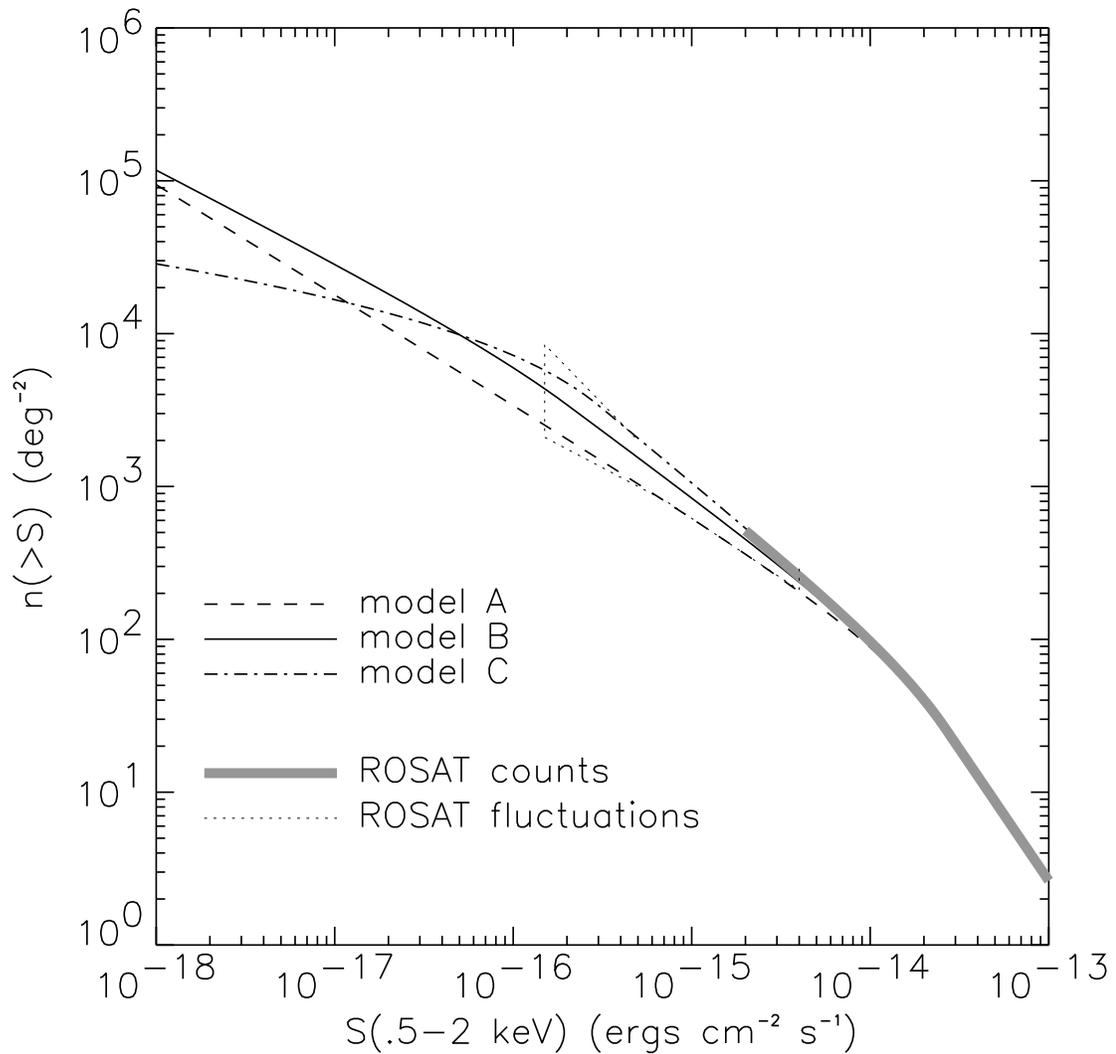}
\caption{Integrated flux distributions for three extrapolated models of
the XRB. Also shown are the ROSAT counts and fluctuation analysis limits
from Hasinger et al. (1993).
\label{fig:xrbmodels}}
\end{figure}

Among the many possible models for the XRB, we consider only three
cases: a model with a moderate slope half way between the fluctuation
analysis limits (model B), and two extreme models with slopes at
$S_{\rm ROSAT}$ just consistent with the fluctuation limits (models A
and C). The second break was set at $S_{23} = 2 \times 10^{-16}$ ergs
cm$^{-2}$ s$^{-1}$, close to the boundary of the fluctuation analysis
limits. These three models are shown in
figure~\ref{fig:xrbmodels}. Their parameter values are given in
table~\ref{tab:xrbmodels} along with those for the observed ROSAT
number-flux relation which was modeled by Hasinger et al. (1993) as
two broken power laws. The ROSAT counts are also displayed in the
figure along with the fluctuation analysis limits. The moderate (B)
and steep (C) models result in 100\% of the XRB being produced by
point sources alone. For the flat model (A), the contribution from
point sources is only 77\%, allowing for a small additional
(e.g. diffuse) component of the XRB.

\placefigure{fig:xrbmodels}

\placetable{tab:xrbmodels}

\section{Effect of Lensing on the X-ray Background}
\label{effect_lensing}
\subsection{General Results}
\label{effect_lensing_general}
We consider a region of the sky where the magnification due to
gravitational lensing has a value $\mu$. The magnification has two
effects on background point sources: their fluxes are
magnified by factor of $\mu$ and their surface number density
is diluted by the same factor. The observed flux and number
density are,
\begin{equation}
S = \mu \hat{S}
\label{eq:s_lensed}
\end{equation}
and 
\begin{equation}
n= \hat{n} / \mu,
\end{equation}
where the unlensed quantities are denoted by a hat. Consequently, the
observed differential number density is
\begin{equation}
\left. \frac{dn}{dS} \right|_{S} = \frac{1}{\mu^2}
  \left. \frac{d\hat{n}}{d\hat{S}} \right|_{S/\mu}.
\label{eq:dnds_mu}
\end{equation}
When the unlensed flux distribution is a power law of the form
$\left. ({d\hat{n}/d\hat{S}}) \right|_{\hat{S}} \propto
\hat{S}^{-\beta}$, the observed counts follow $\left. ({dn}/{dS})
\right|_{S} \propto \mu^{\beta-2} {S}^{-\beta}$.
Thus, the differential counts increase (decrease) as $\mu$ increases, if
$\beta$ is above (below) the critical slope $\beta_{crit} \equiv 2$.

Since we ignore source clustering, the fluctuations in the number
counts are solely due to Poisson statistics. The differential number
of sources in a cell of solid angle $\Omega_c$ is
$\left. ({dN}/{dS})\right|_{S} = \Omega_c
\left. ({dn}/{dS})\right|_{S}$, and its differential variance is
\begin{equation}
\left. \frac{d\sigma^2_N}{dS} \right|_S =
  \Omega_c \left. \frac{dn}{dS} \right|_{S}.
\label{eq:dsign2ds}
\end{equation}
Another useful quantity is the differential XRB intensity 
which is related to the differential counts through
\begin{equation}
\left. \frac{di}{dS} \right|_{S} = S \left. \frac{dn}{dS} \right|_{S}.
\label{eq:dids}
\end{equation}
The differential intensity in a cell is $\left. ({dI}/{dS})
\right|_{S} = \Omega_c \left. ({di}/{dS}) \right|_{S}$, and its
variance is
\begin{equation}
\left. \frac{d\sigma^2_I}{dS} \right|_S =
S^2 \Omega_c \left. \frac{dn}{dS} \right|_S.
\label{eq:dsigi2ds}
\end{equation}

It is often convenient to integrate these quantities between two
observed fluxes, e.g. $n(S_{min},S_{max}) \equiv
\int_{S_{min}}^{S_{max}} dS \left. ({dn}/{dS})\right|_{S}$. The
limits $S_{min}=S_{r}$ and $S_{max}=\infty$ correspond to the resolved
component of the XRB above the detection threshold $S_{r}$, and will be
denoted hereafter as $n(>S_{r})$. The unresolved component corresponds to
the limits $S_{min}=0$ and $S_{max}=S_{r}$, and will be denoted as
$n(<S_{r})$.
\label{integral_notation}

By combining equation~(\ref{eq:dnds_mu}) with
equations~(\ref{eq:dsign2ds})--(\ref{eq:dsigi2ds}), we obtain the
dependence of $n, \sigma_{N}^2, i,$ and $\sigma_{I}^2$ on $\mu$. In
particular, the total integrated intensity
$i(0,\infty)=\int_{0}^{\infty}dS S \mu^{-2}
\left. ({d\hat{n}}/{d\hat{S}})\right|_{S/\mu} = \int_{0}^{\infty}
d\hat{S} \hat{S} \left. ({d\hat{n}}/{d\hat{S}})\right|_{\hat{S}}$, is
invariant under magnification. This results from the 
conservation of the total surface brightness by gravitational lensing
(e.g., Schneider et al. 1992, p. 132).  However, the partial intensity
$i(S_{min},S_{max})$ is {\it not} invariant under magnification,
because the observed flux limits $(S_{min},S_{max})$ correspond to
unlensed fluxes $(\hat{S}_{min}, \hat{S}_{max})$ which depend on
$\mu$.
\label{i_conservation}

It is interesting to note that the total intensity variance
$\sigma_{I}^2(0,\infty)$, does depend on the magnification. Thus,
gravitational lensing does not affect the mean intensity of the
background but changes its angular fluctuations. Note that if
$\beta<3$ (as is the case for counts in Euclidean space, where
$\beta=\frac{5}{2}$), $\sigma_{I}^2$ does not converge as $S
\rightarrow \infty$. In this case, the integrated intensity variance
is only defined with respect to a given upper-limit on the flux.

\subsection{Application to Lensing of Background X-ray Sources}
\label{lensing_application}
\begin{figure}
\plotone{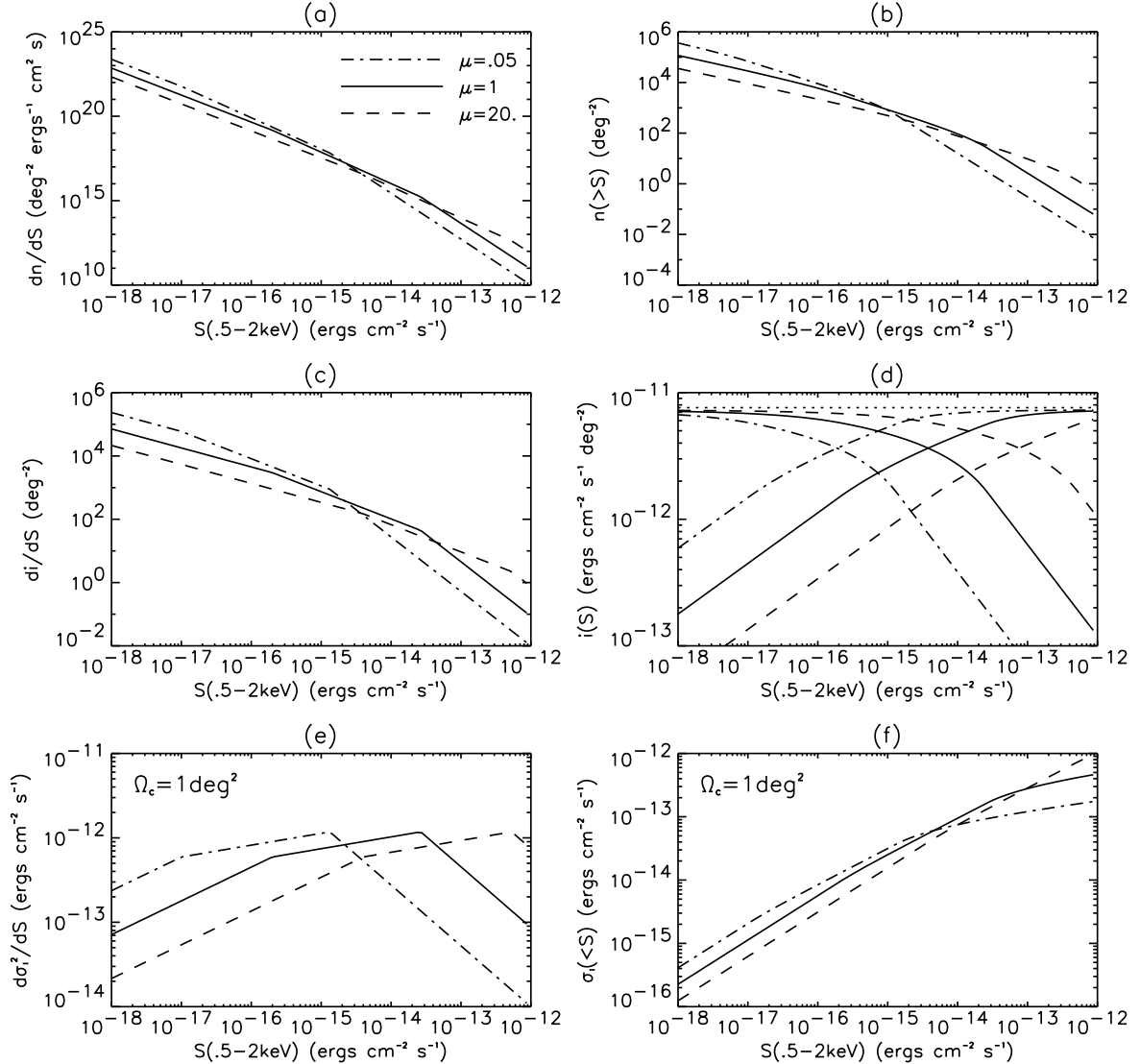}
\caption{Flux distributions for: (a,b) the number density $n$ of X-ray
sources; (c,d) the mean XRB intensity $i$; and, (e,f) the variance
$\sigma_{I}^2$ of the XRB flux $I$ for a cell of solid angle $\Omega_c
= 1$ deg$^2$. The results are for model B of the XRB. We show the
unlensed distributions ($\mu=1$) together with the apparent
distributions for a magnified ($\mu=20.$) and a de-magnified
($\mu=.05$) region of the sky.  Differential and integrated quantities
are shown on the left and right columns, respectively. In panel (b)
the integration limits are from $S$ to $\infty$ (resolved component),
whereas in panel (f) the limits are from 0 to $S$ (unresolved
component).  In panel (d), both the resolved and unresolved
distributions are shown as the decreasing and increasing curves,
respectively.  The invariant total XRB intensity (Hasinger et
al. 1993) is shown as the dotted line on the same panel.
\label{fig:nis_s}}
\end{figure}

We now apply the general relations discussed above to lensing of
background X-ray sources.  The solid line in figure~\ref{fig:nis_s}
shows the unlensed relations ($\mu=1$) for the number density,
intensity, and intensity variance for model B as a function of the
0.5--2 keV flux. The variance of the number density is not plotted
since it is simply proportional to the number density
(cf. Eq.~[\ref{eq:dsign2ds}]).  The columns on the left and right hand
side correspond to the differential and the integrated quantities,
respectively. The resolved and unresolved integrated intensities are
both plotted in panel (d), together with the total XRB intensity
measured by ROSAT (Hasinger et al. 1993) which is shown as the dotted
line. Panel (e) implies that, in this model, most of the XRB
fluctuations originate from sources with fluxes between $\sim10^{-16}$
and $5 \times 10^{-13}$ ergs cm$^{-2}$ s$^{-1}$. For all models (and
for the ROSAT counts; see discussion after
Eq.~[\ref{eq:model-scalings}]), $\beta_1=2.72$, i.e. smaller than 3.
Therefore, the variance of the integrated intensity (cf. panel [f]),
$\sigma_{I}^{2}(<S)$, diverges at bright fluxes but is well defined
for a given upper-limit on the flux.

\placefigure{fig:nis_s}

To illustrate the effect of magnification, figure~\ref{fig:nis_s}
shows the various statistical quantities for $\mu=0.05$ (dot-dashed),
1 (solid), and 20 (dashed).  Note that for $S<S_{*} \approx 10^{-15}$
ergs cm$^{-2}$ s$^{-1}$, the number of resolved sources $n(>S)$
decreases as $\mu$ increases (panel [b]). This so-called
``magnification bias'', is a consequence of the fact that, at faint
fluxes, the differential count slope $\beta$ is smaller than
$\beta_{crit}$.  However, the integrated intensity $i(>S)$ resolved
above the same flux always increases when $\mu$ increases (panel [d]).
Even though the magnification reduces the surface density of faint
resolved sources, these sources appear brighter, and thus their
integrated intensity increases.  The integrated variance
$\sigma_{I}^2(<S)$ of the unresolved intensity decreases when $\mu$
increases for $S<S_{*}$. However, the fluctuation
$\sigma_{I}(<S)/I(<S)$ of the unresolved background (not shown on the
figure) increases with $\mu$.

An inspection of figure~\ref{fig:nis_s} reveals that panel (d) is
qualitatively different from panels (b) and (f) since it does not show
any cross-over between the different magnification lines. In other
words, $i(<S)$ always decreases with $\mu$ for all values of $S$. In
appendix~\ref{app:qualitative_effects}, we show that this is generally
true, to first order in $\mu-1$, for any unlensed number count
relation $\left. {d\hat{n}}/{d\hat{S}} \right|_{\hat{S}}$, provided
that the relevant integrated quantities converge. In the same
appendix, we show that the cross-over flux $S_{*}$ for the weak
lensing equivalent of panel (b) can be conveniently determined by
comparing the solid lines in panels (b) and (c).

In summary, the magnification due to lensing effectively redistributes
the intensity of the XRB into fewer but brighter sources, without
altering the total intensity. For $S<S_{*}$ and for $\mu>1$, the
number of resolved sources is decreased but their integrated intensity
is increased. For the same flux limits, the unresolved XRB appears
dimmer but with larger fluctuations.

\begin{figure}
\plotone{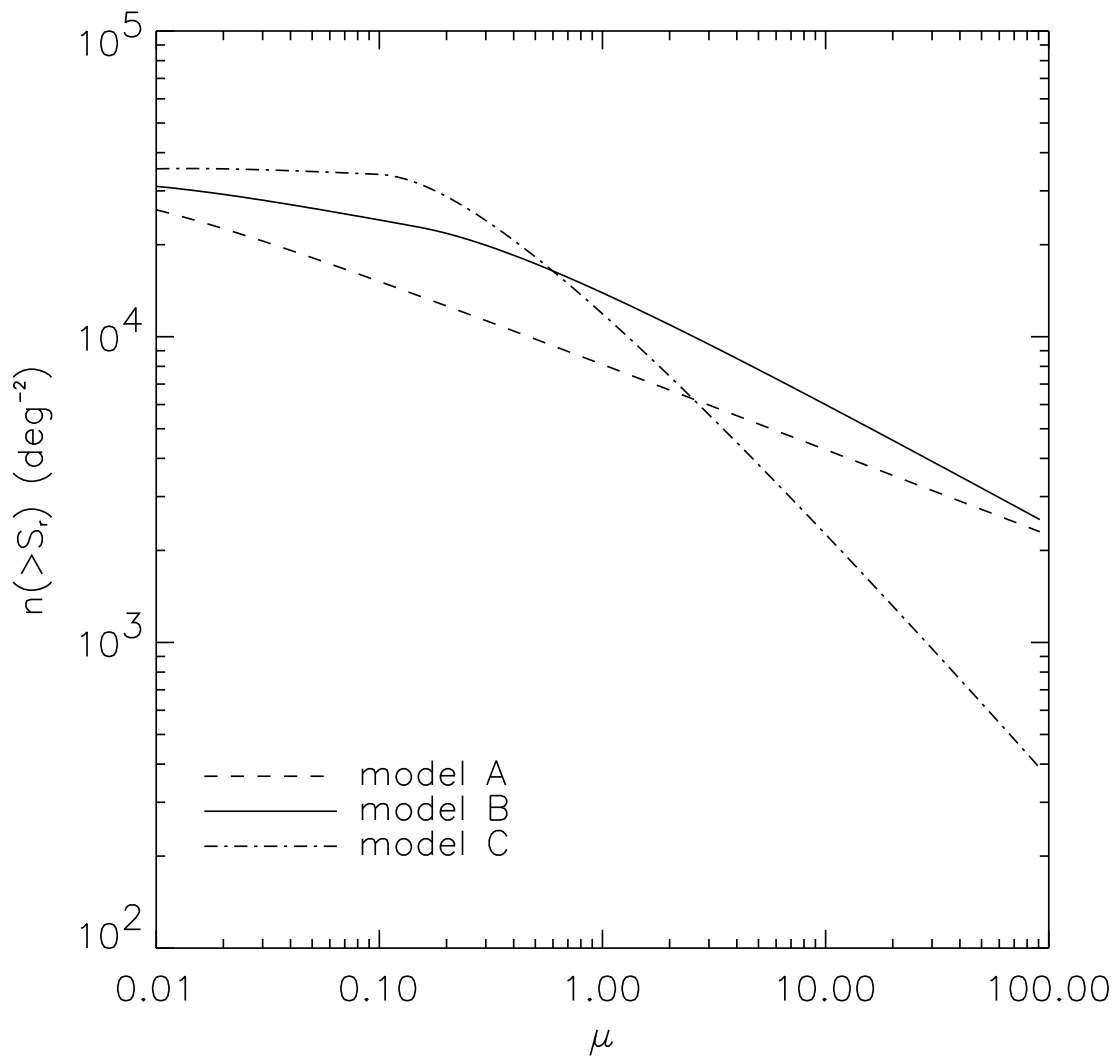}
\caption{Dependence of the resolved source density $n(>S_{r})$ on
the magnification $\mu$ for a detection threshold of
$S_{r}(0.5-2\mbox{keV})= 3 \times 10^{-17}$ ergs cm$^{-2}$ s$^{-1}$.
Results are plotted for each of the three XRB models.
\label{fig:n_mu}}
\end{figure}

We now consider the direct functional dependence of the various
observable quantities on $\mu$. Figure~\ref{fig:n_mu} shows
$n(>S_{r})$ as a function of $\mu$ for all three models. The detection
threshold was fixed to $S_{r}(0.5-2\mbox{keV})=3 \times 10^{-17}$ ergs
cm$^{-2}$ s$^{-1}$. When considering the observability of the effect
in practical cases [see \S\ref{variable_threshold} below], we will
later take full advantage of the radial fall off of the cluster
emission by adopting a variable detection threshold.  Since
$\beta_{3}<\beta_{crit}$ for all three models, $n(>S_{r})$ is a
decreasing function of $\mu$ in all cases. The ``knees'' apparent for
models B and C are due to the power law break at the flux $S_{23}$,
which is shifted across the detection threshold by the
magnification. With $\mu>1$, the relation is steepest for model C, and
flattest for model A. This difference is primarily due to the
different values of $(\beta_{3}-\beta_{crit})$ in these models, and
can thus be used to observationally distinguish between these models.

\placefigure{fig:n_mu}

\begin{figure}
\plotone{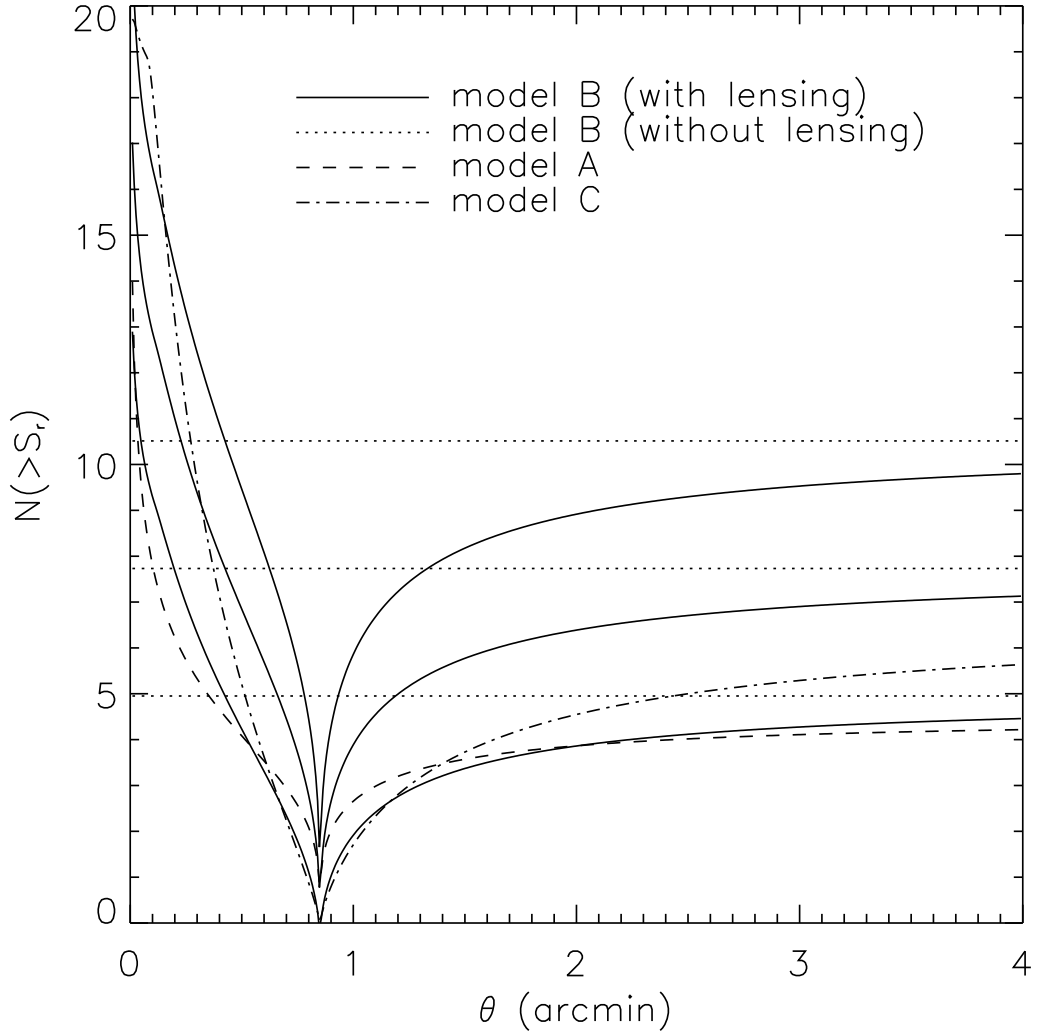}
\caption{Radial dependence of the number of sources with fluxes above
$S_{r}=3 \times 10^{-17}$ ergs cm$^{-2}$ s$^{-1}$ in concentric
annular cells with a solid angle of $\Omega_{c}=2$ arcmin$^{2}$. The
three dotted lines correspond to the unlensed case. The three solid
lines show the effect of lensing by a SIS with an Einstein angle of
$\alpha=0^{\prime}.85$ for model B of the XRB.  The central line in
each set shows the mean number of sources, while the two outer lines
correspond to a single Poisson standard deviation $\sigma_{I}$ away
from the mean. The mean counts for models A and C are indicated by the
dashed and dot-dashed curves, respectively.
\label{fig:n_th_theory}}
\end{figure}

By combining equations~(\ref{eq:mu_theta}) and (\ref{eq:dnds_mu}), one
can derive the dependence of $N, I$ and $\sigma_{I}$ on the angle
$\theta$ relative to the center of a SIS lens. As an example,
figure~\ref{fig:n_th_theory} shows $N(>S_{r})$ as a function of
$\theta$ for model B and for an annular cell size of $\Omega_{c}=2$
arcmin$^2$. The detection threshold was fixed to the value mentioned
above and the Einstein radius $\alpha$ was chosen to be $0^\prime.85$,
as in A1689.  The unlensed case is shown as the dotted lines. In both
cases, the central curve traces the mean count, while the two
neighboring curves correspond to a single Poisson standard deviation
$\sigma_{N}$ about the mean.  The deficiency in the number of resolved
sources close and beyond the Einstein radius is visible in the lensed
case. For comparison, the mean lensed counts for models A and C are
also shown. As expected, model C produces the steepest count profile.
The observability of the source deficit will be discussed in
\S\ref{obs_resolved}.

\placefigure{fig:n_th_theory}

It is instructive to examine the angular dependence of the resolved
source counts at large radii. For $\theta \gg \alpha$,
equation~(\ref{eq:mu_theta}) yields $\mu \approx 1+\alpha/\theta$.
For a sufficiently small detection threshold $S_{r}$, our three XRB
models are dominated by the faint--end counts and thus $N(>S_{r})
\approx \mu^{\beta_3-2} \hat{N}(>S_{r})$, where $\beta_3$ is the power
law index at the faint end.  As a result, $(N - \hat{N}) \propto
\hat{N} \theta^{-1}$. It is convenient to define the signal-to-noise
ratio SNR$_{N}$ relevant for distinguishing lensed from unlensed
source counts as
\begin{equation}
\mbox{SNR}_{N} \equiv \frac{|N - \hat{N}|}{\sqrt{\hat{N}}}.
\label{eq:snrn_def}
\end{equation}
This quantity behaves as SNR$_{N} \propto (\Omega_c)^{\frac{1}{2}}
\theta^{-1}$, where $\Omega_c$ is the solid angle of the annular observing
cell with an angular radius $\theta$. Thus, one can keep SNR$_{N}$
constant by choosing bins with $\Omega_c \propto \theta^2$.  One can
also define the signal-to-noise ratio SNR$_{I}$ for separating the
lensed intensity, $I(<S_{r})$, from the unlensed intensity
$\hat{I}(<S_{r})$ of unresolved sources,
\begin{equation}
\mbox{SNR}_{I} \equiv \frac{|I - \hat{I}|}{\sqrt{\hat{I}}}.
\label{eq:snri_def}
\end{equation}
Through a similar argument, SNR$_{I}$ also maintains constancy for
constant logarithmic bins of solid angle with $\Omega_c \propto \theta^2$.
\label{log_rings}

\subsection{Simulations}
In order to substantiate the above analytic relations and to assess
more realistically whether the lensing signal is detectable, we have
performed numerical simulations of realistic X-ray observations.
Our simulations generate a set of sources with random fluxes
distributed according to the logN-logS relation for one of the XRB
models. We include sources with fluxes in the range of $10^{-20}$ to
$10^{-12}$ ergs cm$^{-2}$ s$^{-1}$ in the 0.5-2 keV band. This range
is sufficiently broad to yield counts and intensities which are
virtually indistinguishable from the ones expected in the full
models. The sources were then assigned random positions in a square
field of size $4' \times 4'$. For the lensed case, we considered a SIS
lens with an Einstein angle $\alpha=0'.85$, similar to that of A1689.
The positions of the lensed images were derived from those of the
unlensed sources by inverting equation~(\ref{eq:lensing}), and their
fluxes were computed using equations~(\ref{eq:mu_theta}) and
(\ref{eq:s_lensed}).

We then simulated observations of the lensed and unlensed fields with
an X-ray imaging instrument for a given exposure time $t_{exp}$.  We
have adopted a Gaussian Point-Spread Function (PSF) with a
one-dimensional standard deviation $\sigma_{psf}=0^{\prime\prime}.21$.
The 0.5--2 keV flux to photon count rate conversion coefficient was
taken to be $c'=3.12 \times 10^{11}$ cts ergs$^{-1}$ cm$^2$ s. These
parameters correspond to the expected performance of the ACIS camera
on board AXAF for 0.2--10 keV observations of sources with a spectral
index $\gamma_{b}=0.4$ (see \S\ref{axaf} below).

\begin{figure}
\plotone{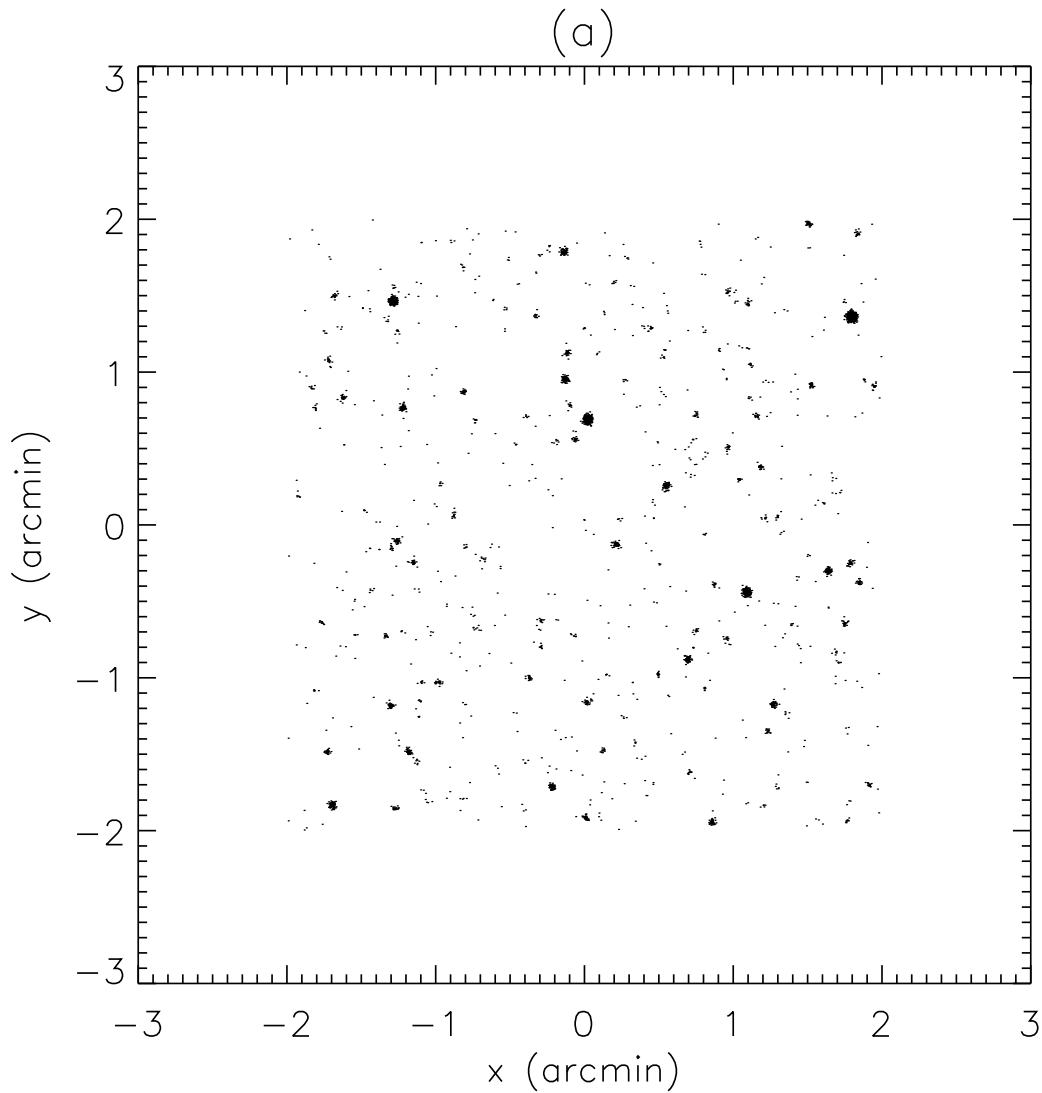}
\caption{Simulated photon maps for model B sources with 0.5-2 keV
fluxes in the range of $10^{-20}$ to $10^{-12}$ ergs cm$^{-2}$
s$^{-1}$.  The detector parameters are those expected for the AXAF-ACIS
camera. However, for clarity we used $\sigma_{psf}$ which is four
times larger than expected. The exposure time is $10^{6}$ sec. The maps
correspond to: (a) unlensed sources; and (b) sources lensed by a SIS
with $\alpha=0'.85$, the Einstein angle of A1689.
\label{fig:pmap}}
\end{figure}

An example of the resulting photon maps for the unlensed and lensed
cases is shown in figure~\ref{fig:pmap}. Here we have chosen the
exposure time to be $t_{exp}=1 \times 10^6$ s, and have adopted model
B for the XRB.  In order to improve the clarity of the picture,
$\sigma_{psf}$ was degraded in this image to a value four times larger
than that quoted in the previous paragraph.  Note that the cluster
emission is {\it not} shown on this image.  The lensed photon map
reveals a faint but noticeable ring close to the Einstein angle where
sources are diluted. The presence of a few bright sources in this ring
keeps its integrated intensity unchanged to within one standard
deviation.

\placefigure{fig:pmap}

\section{Observability}
\label{observability}
In this section, we discuss various strategies that could be adopted to
search for the lensing effect in clusters.  After discussing the
relevant X-ray instrumentation of AXAF in \S\ref{instruments}, we
consider the effect of lensing on the number counts of resolved
sources in \S\ref{obs_resolved}. An alternative approach involves
measuring the unresolved intensity of the XRB below a given detection
threshold, and is considered in \S\ref{obs_unresolved}.  Finally,
\S\ref{obs_other} summarizes other possible observing strategies.

\subsection{X-Ray Instrument}
\label{instruments}
The main limiting factors for observing the lensing effect are the
cluster X-ray emission and the potentially low X-ray source
density. Both of these limitations can be circumvented by using an
instrument with a high angular resolution. Such an instrument would
resolve faint sources even in the presence of the high effective
background resulting from the cluster emission.  The future Advanced
X-ray Astronomy Facility (AXAF) mission (Elvis et al. 1995) is highly
promising in this regard. The AXAF CCD Imaging Spectrometer (ACIS)
camera on board the satellite has a projected PSF with a FWHM of about
$0^{\prime\prime}.5$ (i.e. $\sigma_{psf}=0^{\prime\prime}.21$) and a
pixel size of $0^{\prime\prime}.5$. For our purposes, we will ignore
variations of the PSF across the field.  The effective area of the
instrument is 650 cm$^2$ at 1 keV. It is sensitive in the 0.2-8 keV
band, and its field of view is $16^\prime \times 16^\prime$. The High
Resolution Camera (HRC) in the same mission has a larger field of
view, $32^{\prime} \times 32^{\prime}$, but a smaller effective area,
300 cm$^{2}$ at 1 keV. In this paper, we quote our results for the
ACIS instrument because of its larger effective area.

Of particular interest is the detection capability of the AXAF-ACIS
camera.  In estimating its detection threshold, we consider a source
detection scheme often used in X-ray surveys (e.g., Hamilton et
al. 1991; Hasinger et al. 1993) which consists of counting photons
inside a detection cell. The solid angle $\Omega_{c}$ of the detection
cell is chosen to contain a substantial fraction $f_{p}$ of the total
power of a point source placed at its center. Assuming Poisson
statistics for the photons, the signal-to-noise ratio SNR$_{r}$ for
the resolution of a point source is then
\begin{equation}
\mbox{SNR}_{r}= \sqrt{t_{exp}} \frac{R_{s}}{\sqrt{R_{s}+R_{b}}},
\label{eq:snr_r}
\end{equation}
where $t_{exp}$ is the exposure time, and $R_{s}$ and $R_{b}$ are the count
rates in the detection cell produced by the source and the local
background, respectively. The total background counts are generally the sum
of the internal instrument background, the XRB, and the cluster
emission. The exact count rate of the internal background is not
currently available for the AXAF-ACIS camera, but is likely to be smaller
than the cosmic XRB and therefore much smaller than the cluster emission.
We therefore neglect the internal background in our calculation.

The flux threshold $S_{r}$ is related to $R_{s}$ by $S_{r} = c R_{s}
f_{p}^{-1}$, where $c$ is a flux-to-count-rate coefficient which
depends on the energy band, the instrument effective area, and the
source spectrum. For the AXAF-ACIS camera and for sources with a
spectral index $\gamma_{b}=0.4$ and $N_{H}=1.5 \times 10^{20}$
cm$^{-2}$, the conversion coefficient\footnote[1]{Conversion
coefficients were computed using a version of the PIMMS program which
includes the projected AXAF parameters (Elvis et al. 1995). This
program was kindly provided to us by P. Slane and W. Forman.}  in the
0.2-10 keV band is $c= 7.39 \times 10^{10}$ cts ergs$^{-1}$ cm$^{2}$.
Of interest here is the conversion coefficient from fluxes in the
0.5--2 keV band into the AXAF count rate between 0.2--10 keV, which is
$c'=3.12 \times 10^{11}$ cts ergs$^{-1}$ cm$^{2}$ for the same
parameters.
\label{sr_conversion}

\begin{figure}
\plotone{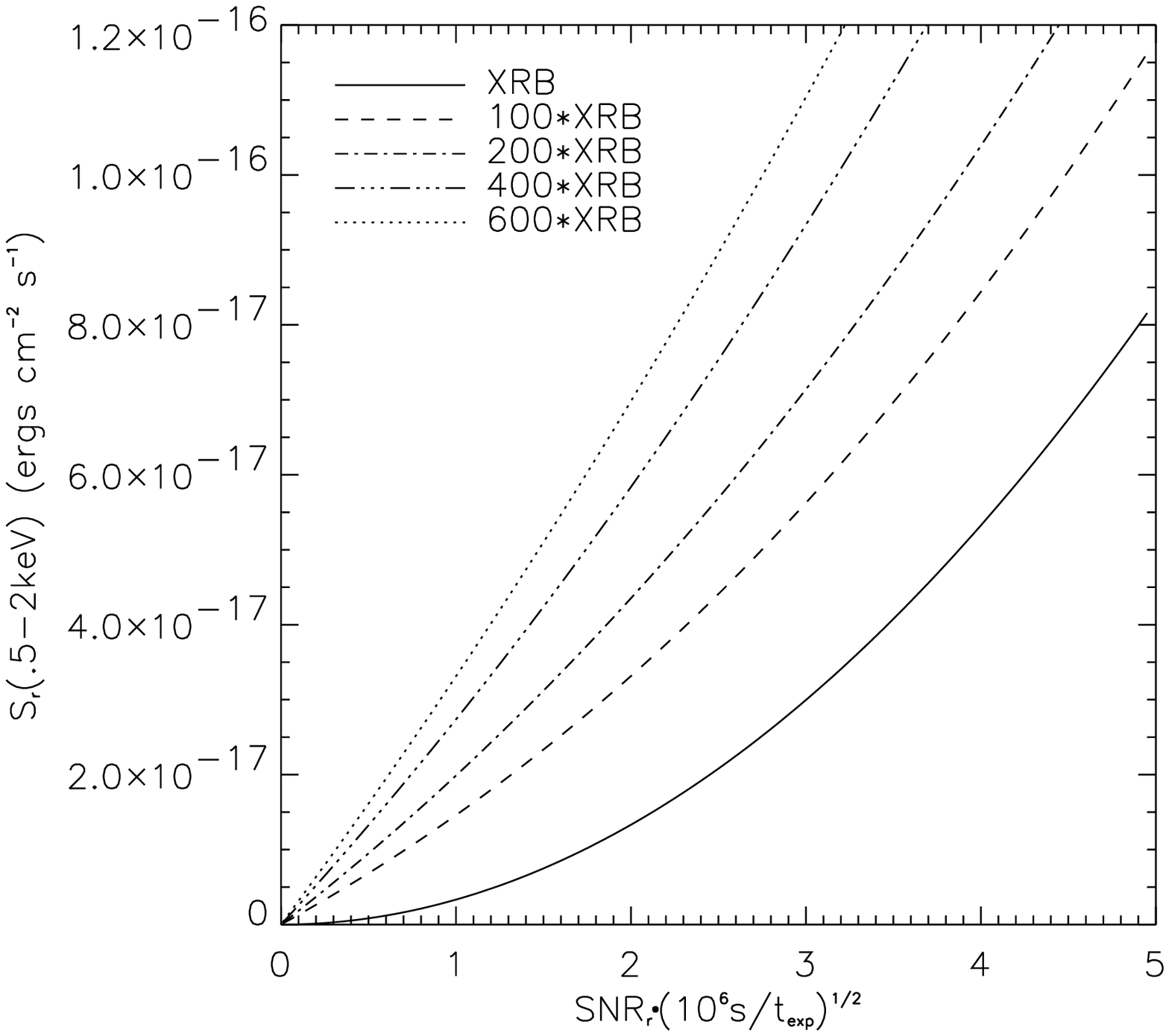}
\caption{Estimated detection capability of the AXAF-ACIS camera.
Detection flux thresholds are plotted as a function of the source
signal-to-noise ratio. The detection is performed in the 0.2--10 keV
band. However, fluxes are quoted in the more familiar 0.5--2 keV
band. The dependence on the exposure time $t_{exp}$ was factored
out. 
The solid line represents an observation in the field, i.e. with
the background equal to the XRB. The other lines correspond to
different background count rates in units of the XRB count rate in the
0.2--10 keV band, as measured by Gendreau et al. (1995).
\label{fig:sn}}
\end{figure}

The resulting flux detection threshold is plotted as a function of the
signal-to-noise ratio in figure~\ref{fig:sn}.  Even though the
detection occurs in the 0.2--10 keV band, the source fluxes are quoted
in the more familiar 0.5--2 keV band. The fraction of the power in a
detection cell was set to $f_{p}=0.96$. This corresponds to the PSF
power contained in $2 \times 2$ ACIS pixels.  The solid line gives the
AXAF detection performance for a background count rate equal to that
of the XRB, i.e.  for an observation in the field. The other curves
correspond to increasing values of the background count rate in units
of the XRB count rate. Note that, due to the degradation of the PSF,
$S_{r}$ could be somewhat larger for sources observed off-axis. The
presence of a high background level degrades the detection
capabilities of the instrument. It is, however, striking that, for a
background level equal to 600 times that of the XRB, the flux
threshold at $2.5\sigma$ is only 4.3 times higher than that in the
field. This relative insensitivity to the background level results
from the high angular resolution of the AXAF telescope. Note that a
zero background level corresponds to a curve which is virtually
indistinguishable from the solid curve in figure~\ref{fig:sn}, which
corresponds to a background level equal to the XRB. Thus, the XRB (or
equivalently, an internal background of comparable magnitude) has a
negligible influence on the detection capabilities of this instrument.

\label{axaf}

\placefigure{fig:sn}

\subsection{Resolved Component}
\label{obs_resolved}
The simplest method to observe the lensing effect is to count the
number of discrete sources at different radii about the
cluster center. This technique is similar to that proposed by
Broadhurst et al. (1995) in the optical band (see also Broadhurst
1995a,b). As shown in \S\ref{lensing_application}, the surface density
of faint sources is expected to be diluted at radii beyond the
Einstein angle of the cluster.

In searching for this dilution of faint sources, one can divide the
field into concentric rings centered about the cluster center, and
count the number of resolved sources in each ring above a given value
of the detection signal-to-noise ratio, SNR$_{r}$. Note that, because
of the variable level of cluster emission, the detection threshold
$S_{r}$ for the fixed value of SNR$_{r}$ depends on position (see
figure~\ref{fig:sn}).  As discussed in \S\ref{log_rings}, it is
convenient to choose the ring areas so as to keep the source count
signal-to-noise ratio SNR$_{N}$ (Eq.~[\ref{eq:snrn_def}])
constant. This is achieved by setting the area of each ring $R_{i}$ to
be $\Omega_{r,i}=b \pi \theta_{i}^2$, where $\theta_{i}$ is the median
radius of $R_{i}$ and $b$ is a dimensionless constant. The resulting
ring radii will thus appear equally-spaced on a logarithmic angular scale.
\label{variable_threshold}

\begin{figure}
\plotone{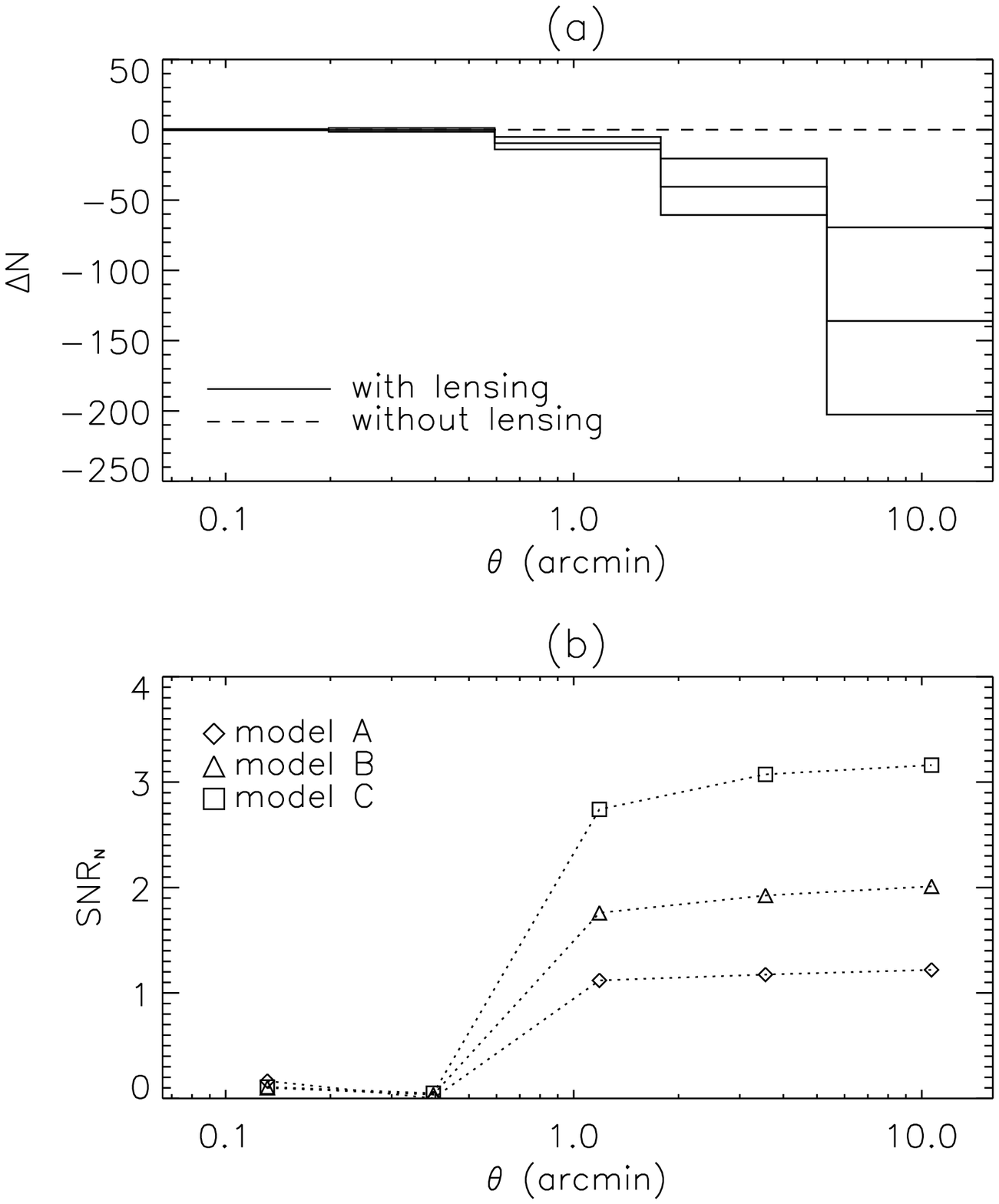}
\caption{Effect of lensing on the resolved source counts for a $10^6$ s
observation of A1689 with the AXAF-ACIS camera. For clarity, the angular
range was extended to $\theta=16'$, i.e. twice the field of view of this
instrument. The source detection threshold was set to
SNR$_{r}=2$. The sources are binned in concentric rings centered about
the cluster center with an area ratio of $b=8$.  Panel (a) shows the
source difference $\Delta N$ between the lensed and unlensed source counts
for model B. The central solid line shows the mean count difference,
whereas the two neighboring lines correspond to a single standard
deviation $\sigma_{N}$ from the mean. The dashed line corresponds
to $\Delta_{N}=0$, i.e. to an observation of the unlensed XRB.
Panel (b) shows the signal-to-noise ratio SNR$_{N}$ for separating the
lensed from the unlensed counts. SNR$_{N}$ is shown for each ring and
for each of the three XRB models.
\label{fig:n_rings}}
\end{figure}

We present our results for the specific case of A1689.
Figure~\ref{fig:n_rings}a shows the expected difference $\Delta N
\equiv N - \hat{N}$ between the lensed and unlensed detected source
counts in each ring. This figure corresponds to a $t_{exp}= 10^6$ s
observation of this cluster with the AXAF-ACIS camera. Since we do not
require a high accuracy in the position and fluxes of the detected sources,
we took SNR$_{r}$ to be as low as 2~.  For the sake of clarity, we show the
source counts from $\theta=0$ to $16'$. This angular range could be covered
by a mosaic of 4 adjacent $16' \times 16'$ ACIS images. In these
calculations, we have adopted model B for the XRB. The lensing signal is
generally maximized for a particular value of the ring area ratio $b$. In
the case under consideration we adopted the optimal value of $b\approx8$.
The central solid line shows the mean count difference, whereas the two
neighboring lines corresponds to a single standard deviation $\sigma_{N}$.
The dashed line corresponds to the expected count difference in the absence
of lensing, i.e. to $\Delta N = 0$.

The deficit of lensed counts as compared to the unlensed counts is
visible for the three outer rings with $\theta \gtrsim 0^\prime.6$.
The cluster emission precludes the possibility of observing the
lensing effect at smaller radii. Figure~\ref{fig:n_rings}b shows the
signal-to-noise ratio SNR$_{N}$ for the difference between the lensed
and the unlensed counts (Eq.~[\ref{eq:snrn_def}]) in each ring. For
model B, the three outer rings have values of SNR$_{N}$ of 1.8, 1.9,
and 2.0, from low to high values of $\theta$. In
appendix~\ref{app:stat_n}, we present a method for estimating the
significance of the combined signal from several rings based on
$\chi^2$ statistics. Using this method, we find that the combined
source counts in the outer three rings at hand can be distinguished
from the unlensed counts at the $2.9\sigma$ confidence level.

Figure~\ref{fig:n_rings}b also shows the ring statistics for models A and
C. Because of its flat faint logN-logS relation, model C produces the
largest signal. The combined significance for the outer three rings is 1.8
and 4.8 for models A and C, respectively. The large difference between the
combined significance of each model illustrates the sensitivity of the
lensing effect to the XRB model. 

If only one ACIS pointing is available, one can only count sources out
to $\theta=8'$. Since part of the most outer ring in
figure~\ref{fig:n_rings} would then be lost, the lensing signal would
be somewhat reduced. For the same conditions as quoted above but with
$b=15$, only two rings then have a significant lensing signal. The
combined significance of these two outermost rings is 1.7, 2.7, and
$4.2\sigma$, for models A, B, and C, respectively.

\placefigure{fig:n_rings}

\begin{figure}
\plotone{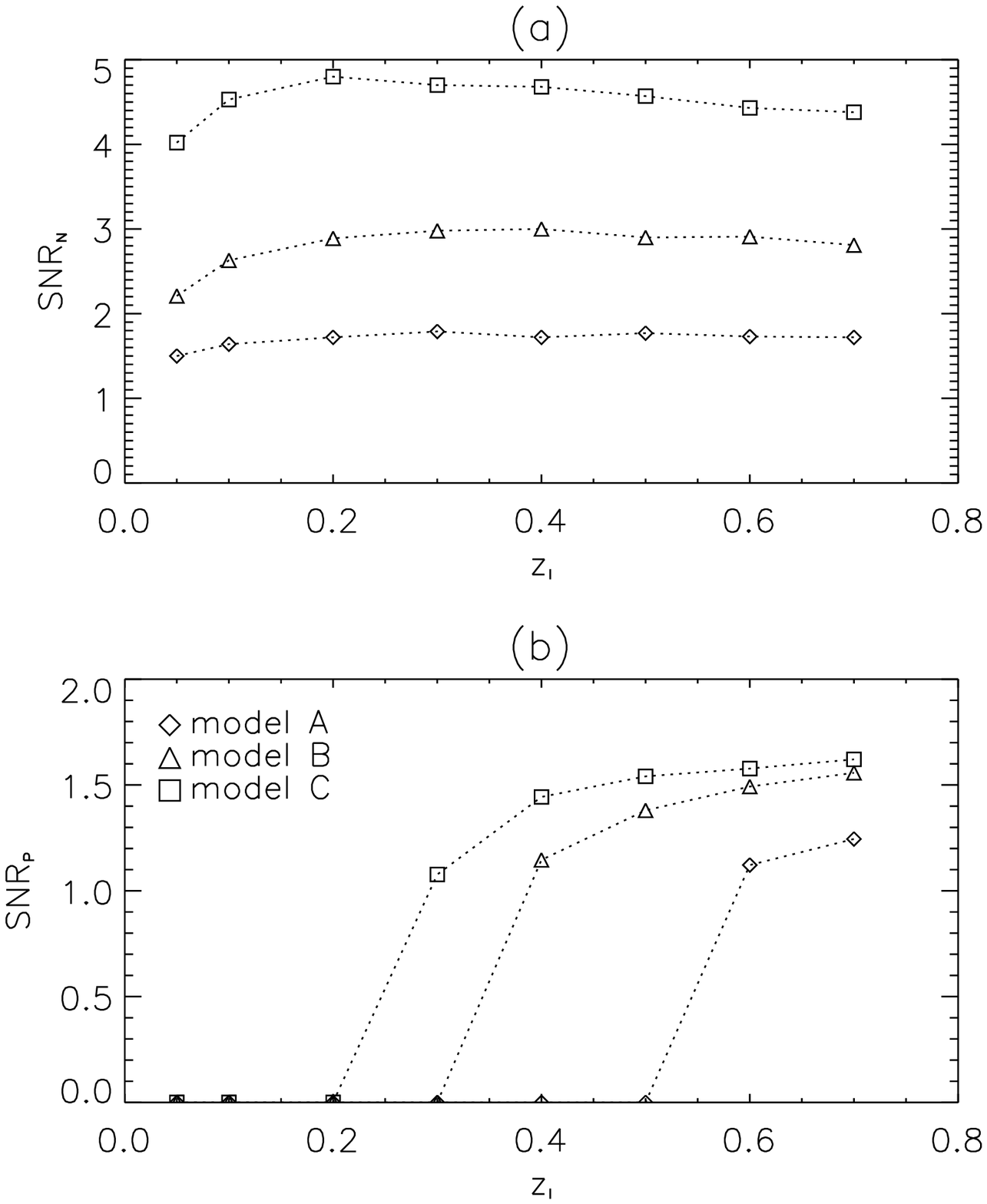}
\caption{Dependence of the lensing signal on the cluster redshift.
Several values of the redshift $z_{l}$ of a displaced version of
A1689 are considered for each of the XRB models, and for a $10^6$ s 
observation with the AXAF-ACIS camera. For clarity, the 
angular range was extended to $\theta=16'$, i.e. to twice the field
of view of this instrument. Panel (a) shows
the combined SNR$_{N}$ for resolved source counts expected
for SNR$_{r}=2$ and $b=9$. Panel (b) shows
the combined SNR$_{P}$ of the photon counts for the unresolved
intensity. We chose $b=3$ and set the detection flux
threshold to $S_{r}(0.5-2\mbox{keV})=1, 6, 30 \times 10^{-17}$
ergs cm$^{-2}$ s$^{-1}$ for model A, B and C, respectively.
\label{fig:snrnp_z}}
\end{figure}

We next examine the redshift dependence of the lensing signal by
considering displaced versions of A1689 (see \S\ref{a1689_displaced}).
Figure~\ref{fig:snrnp_z}a shows the combined SNR$_{N}$ as a function
of the hypothetical cluster redshift $z_{l}$ for each of the XRB
models. The figure corresponds to a $t_{exp}=10^6$ s observation with
the AXAF-ACIS camera out to $\theta=16^{\prime}$ with SNR$_{r}=2$ and
$b=9$. All rings with an individual SNR$_{N}$ greater than unity
were included in the computation of the combined SNR$_{N}$. Notice
that SNR$_{N}$ is rather insensitive to the lens redshift in the range
for $0.15\la z_l\la0.6$.  As the redshift of the cluster is increased,
its Einstein angle decreases but so does the angular radius and
surface brightness of its X-ray core. These competing effects tend to
cancel each other.  One can nevertheless notice that SNR$_{N}$ peaks
at a value of $z_{l}$ which depends on the XRB model; the peak value
is reached at $z_l \approx 0.2$ for model C, and at $z_l \approx 0.4$
for model B. In model C, the lensing signal is stronger and thus the
cluster emission is not as limiting as in the other models.  As a
result, it is advantageous in this model to position the cluster lens
closer, thereby increasing the Einstein angle at the expense of
stronger cluster emission. Note that if the faint background X-ray
sources have a mean redshift substantially smaller than 1, then a
non-negligible fraction of the XRB might be emitted in front of the
distant clusters. In that case, figure~\ref{fig:snrnp_z}a would need
to be corrected for the reduction in SNR$_{N}$ at high values of
$z_{l}$.

\placefigure{fig:snrnp_z}

Even though the detection of lensing through the change in the number
counts of resolved X-ray sources requires a long exposure time with
AXAF ($t_{exp} \sim 10^6$ s), the significance level of the signal is
between 2 and $4\sigma$ in the case of A1689.  The lensing effect can
be used to probe the potential of rich clusters, especially at large
radii. Since the amplitude of the effect is sensitive to the XRB
model, a detection would constrain the faint end of the logN-logS
relation for the background X-ray sources. The lensing signature on
the resolved source counts is rather insensitive to the cluster
redshift, and so many clusters with redshifts in the range $0.1\la
z_{l}\la 0.6$ could have detectable signatures.  Future systematic
surveys of the lensing properties of X-ray clusters (see, e.g. Le
F\`{e}vre et al. 1994) could be used to optimize the selection of
cluster lenses for this study.

\subsection{Unresolved Component}
\label{obs_unresolved}
\subsubsection{Strategy}
Because of the conservation of the total intensity by gravitational
lensing (see \S\ref{i_conservation}), the resolved and unresolved
background intensities are equivalent gauges of the lensing
signal. However, in the presence of the strong cluster emission, the
observational methods used to detect resolved point sources are
significantly different from those used to measure unresolved
intensities. It is therefore useful to examine the possibility of using
the unresolved intensity as a complementary method for confirming the
existence of the lensing effect.

The unresolved component of the XRB is contaminated by the cluster
emission, whose radial profile is unknown {\it a priori}. However, at
large radii, the cluster emission typically falls off rapidly as a
power-law, $i_c\propto\theta^{-\beta_{x}}$, with $\beta_{x} \approx
5.19$ for the analytical fit used to model A1689
(Eq.~[\ref{eq:i_daines}]), or $\beta_{x} \approx 3$ for the King model
(King 1962; Sarazin 1988). On the other hand, as far as the isothermal
mass distribution applies, the signal-to-noise ratio of the lensing
effect on the unresolved intensity falls--off only as $\theta^{-1}$
(Eq.~[\ref{eq:snri_def}]).  The lensing effect could therefore
dominate over the cluster emission at sufficiently large radii. As
discussed in \S\ref{lensing_application} (see also
appendix~\ref{app:qualitative_effects}), lensing tends to reduce the
unresolved intensity of the XRB. Consequently, the outer halo of a
rich cluster might show an annulus where the diffuse X-ray
intensity is lower than the sky-averaged intensity of the XRB, after
the removal of resolved sources. In this sub-section, we will
examine the detectability of this annulus, which is generic to
lensing.

For this purpose, we divide, as before, the field into
logarithmically-spaced concentric rings; this binning scheme
keeps the signal-to-noise ratio of the unresolved intensity SNR$_{I}$
(Eq.~[\ref{eq:snri_def}]) constant at large radii.  After removing all
the point sources with fluxes above a given flux detection threshold
$S_{r}$, the unresolved background intensity can be measured in each
ring. In a given ring, the total X-ray intensity is the sum of the XRB
and the cluster emission intensities, $I_{tot}=I_{b}+I_{c}$. Here, we
define the XRB intensity as $I_{b}(<S_{r})$ in the notation of
\S\ref{integral_notation}. As noted in \S\ref{axaf}, the AXAF
instrumental background is small compared to the cluster flux and can
therefore be neglected.

Let us consider a ring $R$ of solid angle $\Omega_{c}$ and with a mean
angular radius $\theta$. Typically, we consider a solid angle $\Omega_{c}
\ga 400$ arcsec$^2$, which is much larger than the PSF ($\pi \sigma_{psf}^2
\approx 0.14$ arcsec$^2$), and so we ignore the latter.
The total photon counts in $R$ is then $P_{tot}=P_{b}+ P_{c}$ with
\begin{equation}
P_{q}(\theta) \approx t_{exp} c_{q} \int_{R} d\Omega^{\prime}
i_{q}(\theta^{\prime})
\end{equation}
where $q=b,c$, $c_{q}$ are the appropriate flux-to-count rate
conversion coefficients described in \S\ref{sr_conversion}, and the
integration extends over the solid angle of the ring $R$.

The variance of the photon counts $\sigma^{2}_{P,tot}$ can also be
decomposed as $\sigma^{2}_{P,tot}=\sigma^{2}_{P,b}+\sigma^{2}_{P,c}$.
Two effects contribute to $\sigma^{2}_{P,b}$: photon statistics and
the intrinsic XRB fluctuations.  These two effects are generally
correlated since in a region of small intrinsic intensity, $P_{b}$ is
small and thus the Poisson variance in the number of photons is small.
However, for the long exposure time considered in this work, $P_{b}$
is very large, and so this correlation can be neglected.  The variance
of the XRB photon counts in $R$ is then
\begin{equation}
\sigma^{2}_{P,b}(\theta) \approx P_{b} + 
\left( t_{exp} c_{b} \right)^2 
\int_{R} d\Omega^{\prime} \sigma_{i,b}^2(\theta^{\prime}),
\label{eq:sigma2_pb}
\end{equation}
where the first term is due to photon statistics, and $\sigma_{i,b}^2
\equiv \int_{0}^{S_{r}} dS S^{2} \left. ({dn}/{dS})\right|_{S}$ is
the intrinsic variance of the unresolved background intensity (see
Eq.~[\ref{eq:dsigi2ds}]).  The variance in the cluster photon counts,
$\sigma^{2}_{P,c}$, can also be decomposed into intrinsic and photon
statistics terms. Given the strong cluster intensity and the geometry
of the rings, the intrinsic term is probably small compared to the
photon statistics term. We thus neglect the former term and
approximate $\sigma^{2}_{P,c} \approx P_{c}$.

In order to characterize the strength of 
the lensing effect, we consider the difference $\Delta P
\equiv P_{tot}-\hat{P}_{b}$, between the total number of counts observed
($P_{tot}$) and the expected number for the unlensed XRB counts
($\hat{P}_b$).  The unlensed counts, $\hat{P}_b$, can be independently
determined through an average over the entire sky, or through measurements
of the unresolved intensity of the XRB in control regions which are far
from any cluster. The lensing effect is expected to produce negative values
of $\Delta P$. In order to assess the significance of this effect, we define
the signal-to-noise ratio SNR$_{P}$ for detecting a deficit in the photon
counts as
\begin{equation}
\mbox{SNR}_{P} \equiv \frac{P_{tot}-\hat{P}_{b}}{\sigma_{P,tot}}
= \frac{\Delta P}{\sigma_{P,tot}}.
\end{equation}

\subsubsection{Results}
Figure~\ref{fig:i_rings}a shows $\Delta P$ as a function of $\theta$
for model B. The count differences are shown for an AXAF-ACIS
observation with $t_{exp}=10^{6}$ s of A1689 displaced to $z_l=0.6$.
The ring area ratio was set to $b=0.8$.  For clarity, this figure
includes angles in the range $\theta=1$--$32^\prime$, implicitly requiring a
mosaic of sixteen adjacent AXAF pointings. The detection threshold was set
to $S_{r}(0.5-2\mbox{keV})=6 \times 10^{-17}$ ergs cm$^{-2}$ s$^{-1}$.
This threshold was chosen to optimize the lensing effect without
removing too large a fraction of the XRB. The central solid line
corresponds to the mean value of $\Delta P$. The two extreme solid
lines correspond to 
one standard deviation $\sigma_{P,tot}$ away from the mean.  The
dashed line corresponds to $\Delta P=0$, i.e. to an observation of the
unresolved XRB intensity outside the cluster.

\placefigure{fig:i_rings}

The positive values of $\Delta P$ for $\theta \la 10^\prime$ are, of
course, due to the cluster emission. Lensing causes $\Delta P$ to take
negative values for $\theta \ga 10^\prime$, as long as the isothermal mass
profile of the cluster extends out to these scales. As expected, the
deficit in the unresolved XRB intensity only dominates at large angles,
where the cluster emission is sufficiently weak.

Figure~\ref{fig:i_rings}b shows SNR$_{P}$ for each ring and for each
XRB model.  In rings with $\Delta P > 0$, SNR$_{P}$ was set to zero. For
model B, the signal in the three outermost rings (taken separately) has a
significance of SNR$_{P}=$1.2, 1.4 and $1.4\sigma$, respectively.  In
appendix~\ref{app:stat_p}, we describe a method for estimating the
combined significance of the photon counts in several rings.  Using
this method, we obtain a combined significance of $2.0\sigma$ for the
above three rings. The corresponding combined significances for models
A and C are 1.79 and $2.06\sigma$, with $S_r(0.5-2\mbox{keV})=1 \times
10^{-17}$ and $3 \times 10^{-16}$ ergs cm$^{-2}$ s$^{-1}$,
respectively. The minimum radius, $\theta_{\rm def}$, at which an intensity
deficit is present (i.e. outside which $\Delta P<0$), is smallest in model C.

We can again study the redshift dependence of the lensing effect by
considering different displacement redshifts $z_{l}$ for A1689 (see
\S\ref{a1689_displaced}).  Figure~\ref{fig:snrnp_z}b shows the
combined SNR$_{P}$ as a function of $z_{l}$ for each of the XRB
models. This figure corresponds to a $t_{exp}=1 \times 10^{6}$ s
observation with the AXAF-ACIS camera and $b=0.8$.  In this plot the
angles were restricted to the more realistic range of
$0<\theta<16^\prime$, which corresponds to a mosaic of four adjacent
AXAF-ACIS fields. The flux threshold was set to
$S_{r}(0.5-2\mbox{keV})= 1, 6, 30 \times 10^{-17}$ ergs cm$^{-2}$
s$^{-1}$ for model A, B, and C, respectively.  One readily notices that,
for each model, SNR$_{P}$ is equal to zero up to a redshift
$z_{l,{\rm def}}$ beyond which it increases until it reaches a plateau.
The value of $z_{l,{\rm def}}$ corresponds to the redshift at which
$\theta_{\rm def}$ enters the field-of-view. At the actual redshift
of A1689, none of the models produce a depletion. However, at $z_{l}=0.6$,
the depletion is significant at the 1.1, 1.5 and $1.6\sigma$ level,
for model A, B, and C, respectively. Note that these results
are contingent on the validity of the isothermal sphere approximation
out to the observed radii. 

We conclude that it is more difficult to detect the lensing signature
on the unresolved component of the XRB than it is in the resolved
component case.  However, it is qualitatively interesting that lensing
could create an annulus around rich clusters where the diffuse
unresolved X-ray intensity is smaller than its sky-averaged value. For
clusters like A1689 observed with four adjacent AXAF-ACIS $10^6$s
pointings, the depletion is present only for clusters with $z \ga 0.5$
but with a low detection significance ($\sim 1-1.6\sigma$). The
depletion would be much easier to detect behind clusters which are
X-ray faint, such as hypothetical ``dark'' clusters with anomalously
low gas fractions.

\subsection{Other Observing Strategies}
\label{obs_other}
Another potential method for observing the lensing effect is to
consider the angular fluctuations of the unresolved XRB. As shown in
\S\ref{effect_lensing}, magnification increases the fluctuations in
the unresolved component of the XRB for $S<S_{*}$. Quantitatively, one
could measure the unresolved integrated variance $\sigma_{I}^2(<S_r)$
in concentric rings and attempt to detect an enhancement of the
fluctuation $\sigma_{I}/I$ close to the Einstein radius. We leave the
study of this approach to future work.  A similar method using the
auto-correlation function of the extragalactic background light was
recently proposed for the optical band (Waerbeke et al. 1996; Waerbeke
\& Mellier 1996).

In this paper, we have only considered the detection of the lensing effect
behind a single cluster, but an alternative approach could consist in
combining X-ray observations of several clusters. One of the major limiting
factors in detecting the lensing signal is the low number of resolved
sources behind a single cluster. By stacking X-ray images of several
clusters together, the effective number of resolved sources and the
significance of the lensing signal would be increased. We have seen in
\S\ref{obs_resolved}, that the lensing signal for the resolved source
counts is fairly insensitive to the cluster redshift. Therefore, any
sufficiently massive cluster with $z_{l}$ between 0.1 and 0.6 could be used
for this purpose. The significance of the signal could be maximized by
rescaling the angular size of each cluster in units of its Einstein angle
based on its X-ray temperature.

The same approach can, in fact, be carried one step further and
applied to an all-sky survey which includes a large number of clusters
and background sources.  As shown in
appendix~\ref{app:qualitative_effects}, the intensity of the
unresolved component always decreases due to lensing.  The
cross-correlation between the unresolved XRB intensity and cluster
positions on the sky is therefore expected to be reduced at large
separations ($\sim10^{\prime}$--$1^\circ$) due to lensing, while being
enhanced at small separations due to X-ray emission by the clusters. A
measurement of the cross-correlation function between Abell clusters
and the ROSAT All-Sky Survey intensity was recently performed by
Soltan et al. (1996). In a companion paper (Refregier \& Loeb 1996b),
we plan to examine the effect of lensing on such measurements.

\label{arcs}
Until now, we have modeled X-ray sources as point sources. However, as
discussed in \S\ref{galaxies}, some of the faint X-ray sources could be
galaxies with extended emission and could thus produce lensed arcs. In order
to derive a rough estimate for the probability of detecting an arc in
A1689, we assume that a fraction $f_{e}$ of all X-ray sources are extended.
We take the X-ray intensity of each extended source to be a disk with an
intrinsic angular radius $\hat{\theta}_{e}$. Because of gravitational
lensing, the apparent solid angle of a disk $\Omega_{e}$ is stretched by
the magnification factor, i.e. $\Omega_{e}= \mu \hat{\Omega}_{e}$, where
$\hat{\Omega}_{e} \equiv \pi \hat{\theta}_{e}^{2}$ is the unlensed solid
angle. The source intensity is unchanged while the flux follows
equation~(\ref{eq:s_lensed}). We estimate the number $N_{arcs}$ of
observable arcs to be the number of extended sources which are stretched by
a factor larger than a given magnification, $\mu_{min}$, i.e.
\begin{equation}
N_{arcs} \approx f_{e} \int_{\mu>\mu_{min}} \!\!\!\!\!\!\!\!\!\!\!\!
d\Omega \: n(>\!\!S_{r}(\theta)),
\end{equation}
where the integration is over the region of the sky (typically a ring
centered on the Einstein angle) in which $\mu>\mu_{min}$.  After
choosing a detection threshold SNR$_{r}$, the flux limit $S_{r}$ can
be determined by inverting equation~(\ref{eq:snr_r}). Because we are
now considering extended sources, the solid angle of the cell used for
detecting sources must be set to $\Omega_{e}$ to encompass the total
flux of the source.
The value of $S_{r}$ depends on position because of the cluster
emission.

\begin{figure}
\plotone{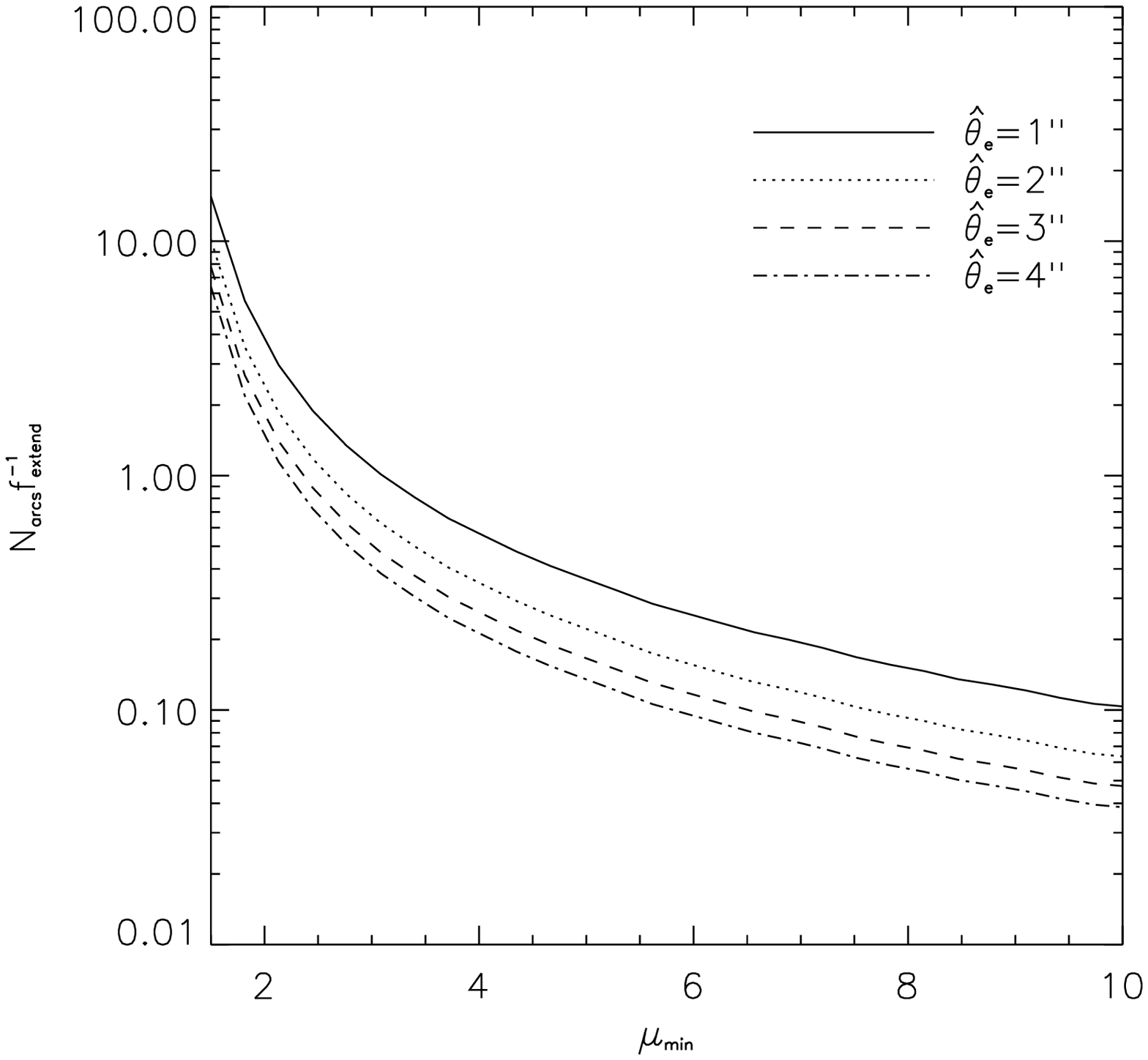}
\caption{Expected number of arcs $N_{arcs}$ as a function of the
minimum magnification (or stretch) factor $\mu_{min}$ for several
intrinsic source sizes $\hat{\theta}_{e}$. The fraction $f_{e}$ of
extended sources in the population of X-ray sources was factored
out on the vertical axis. 
The results correspond to a $10^6$ s observation of A1689 with
the AXAF-ACIS camera, and a source detection threshold of
SNR$_{r}=3$. We assume model B for the XRB.
\label{fig:narcs}}
\end{figure}

Figure~\ref{fig:narcs} shows the resulting expected number of arcs as
a function of $\mu_{min}$ for different values of
$\hat{\theta}_{e}$. The figure corresponds to a $10^{6}$ s observation
of A1689 with the AXAF-ACIS camera. In this plot, we assume a detection
threshold of SNR$_{r}=3$ and adopt model B for the XRB. At a given
value of $\mu_{min}$, the number of arcs which can be resolved is
smaller for intrinsically larger sources because of the cluster
emission. Note, however, that such sources produce larger arcs for a
given value of $\mu_{min}$. For $\hat{\theta}_{e}=1^{\prime\prime}$
and for $\mu_{min}=3$, the expected number of arcs is comparable to
$f_{e}$.

\placefigure{fig:narcs}

\section{Conclusions}
\label{conclusion}
In this work we have analyzed the gravitational lensing signature of a
single rich cluster of galaxies on the XRB. We have found that near
and outside the Einstein angle of the cluster (i.e. for $\theta \ga
1^{\prime}$), lensing results in a deficit of resolved X-ray sources
with fluxes $\ga 3\times10^{-17}~{\rm ergs~cm^{-2}~s^{-1}}$. In the
same region, the corresponding unresolved component of the XRB appears
dimmer and more anisotropic. Although the lensing signal peaks near
the Einstein angle of the cluster (cf.  Fig.~\ref{fig:n_th_theory}),
its detection is easier in the outer regions where the cluster
emission is negligible.

For concreteness, we have considered future observations of the
cluster A1689 with the forthcoming AXAF-ACIS camera. After a $10^6$
second exposure on this instrument, a detection of the lensing
signature imprinted on the resolved background source counts should be
difficult but possible out to $8^\prime$ from the center of
A1689 with a significance level of $\sim 2$--$4\sigma$, depending on
the choice of the extrapolated flux distribution of faint X-ray
sources (see Fig.~\ref{fig:n_rings}). 
Far from the cluster core ($\theta \ga 10'$), the deficit in the
unresolved intensity of the XRB due to lensing may dominate over the
intensity excess due to the cluster emission.  For X-ray bright
clusters, it is, however, difficult to detect this deficit
(cf. Figs.~\ref{fig:snrnp_z} and \ref{fig:i_rings}).

The main factors affecting the strength of the lensing signal are the size
of the Einstein angle and the intensity of the cluster emission. Massive
clusters typically have large Einstein radii, but also tend to have high
X-ray luminosities (see, e.g. Sarazin 1988).  The lensing signature on the
resolved source counts is almost insensitive to the lens redshift in the
range $0.1\la z_l\la 0.6$ (cf. Fig.~\ref{fig:snrnp_z}). Existing surveys of
the X-ray properties of clusters (eg. Ebeling et al. 1996) can be used, in
conjunction with simplifying assumptions (such as spherical symmetry,
relation between velocity dispersion X-ray temperature), to select the most
promising clusters for this study.  Future systematic surveys of the
lensing properties of X-ray clusters (see, e.g. Le F\`{e}vre et al.  1994)
will provide increased confidence in the selection of these clusters.

Currently, little is known observationally about the mass distribution
on scales $\sim 10^\prime$ around clusters.  Far from the cluster
center, the lensing signal-to-noise ratio behaves as $\theta^{-1}$ and
thus falls--off more slowly than the cluster free-free intensity,
which typically decreases at least as rapidly as $\theta^{-3}$.  X-ray
telescopes with a large field of view could therefore probe the mass
distribution as far as the virialization boundary of a cluster, well
outside the regime where the X-ray emission from the cluster is
detectable.

Because the magnification bias due to lensing is sensitive to
the faint end slope of the number-flux relation of X-ray sources, the
amplitude of the lensing signal can constrain different models of the
extrapolations of this relation (cf. Figs.~\ref{fig:xrbmodels} and
\ref{fig:n_rings}). Deep X-ray imaging of cluster fields may 
be used in this way to shed more light on the origin of the XRB.

The nature of the faint population of extragalactic X-ray sources is
still a matter of debate.  In most of this work, we have assumed that
all sources are high--redshift AGN and would thus appear
point-like. However, it is possible that a significant fraction,
$f_{e}$, of them are associated with extended emission from distant
galaxies (Jones et al. 1995; Boyle et al. 1995; Carballo et al.  1995;
Griffiths et al. 1996; however see Hasinger 1996a,b; Refregier et al.
1996). The existence of extended sources behind the lensing cluster
can be tested by a search for X-ray arcs or arclets. For a $10^{6}$
second observation of A1689 with AXAF-ACIS, the number of arclets
which are stretched by a factor $\ga 3$ and are detectable at the
$3\sigma$ level, is roughly comparable to $f_{e}$ for intrinsic source
sizes $\sim 1^{\prime\prime}$ (see Fig.~\ref{fig:narcs}).  The
statistical detection of weak distortions to the intrinsic
ellipticities of sources is routinely done in optical studies of weak
lensing (Schneider 1995; Kaiser 1995; and Narayan \& Bartelmann 1996),
and could also be extended, although with smaller statistics, to the
X-ray band. By observing an ensemble of lensing clusters, it may
therefore be possible to calibrate the contribution of
narrow-emission-line galaxies at high redshift to the XRB.

In the future, we plan to extend the present single cluster analysis
and include the effect of lensing by an ensemble of clusters
(Refregier \& Loeb 1996b). The effect of lensing on the
cross-correlation between clusters and the unresolved intensity of the
XRB is particularly interesting in light of a recent measurement of
the correlation between Abell clusters and the ROSAT All-Sky Survey
(Soltan et al. 1996). Lensing reduces the intensity of the unresolved
background (cf. appendix~\ref{app:qualitative_effects}), and thus
provides a negative contribution to the cross-correlation signal at
large angular separations ($\sim10^{\prime}$--$1^\circ$), where the
cluster emission is negligible.

\acknowledgements We would first like to thank D.J. Helfand for many
useful discussions and suggestions.  We are grateful to W. Forman and P.
Slane for informing us about the projected performance of AXAF and for
providing a version of PIMMS which includes the AXAF parameters. We also
thank M. Bartlemann for insightful comments, P. Fischer for suggesting to
use A1689 as a test case for this study, and the editor E. Wright for a 
particularly helpful correspondence. This work was supported by the grant
NAGW2507 from the NASA LTSA program (for AR) and by the NASA ATP grant
NAG5-3085 (for AL).

\mbox{}  
\begin{center}
{\bf Appendix}
\end{center}
\appendix
\section{Qualitative Effects of Magnification}
\label{app:qualitative_effects}
In this appendix, we address the question whether the resolved source
counts $n(>S)$ and the unresolved intensity $i(<S)$ decrease or
increase as the magnification $\mu$ increases.  The results are not
only applicable to the XRB, but also hold for any background
consisting of a collection of randomly distributed point sources which
are sufficiently distant from the lens.

For the purpose of this discussion, it is convenient to rename the
unlensed differential counts as $f(\hat{S}) \equiv \left.
({d\hat{n}}/{d\hat{S}}) \right|_{\hat{S}}$ (see notation in
\S\ref{effect_lensing_general}). We also rewrite the magnification as
$\mu=1+\epsilon$, and consider the weak lensing regime in the outer
parts of the lensing cluster where $\epsilon\ll1$.

\subsection{Resolved Source Counts}
The resolved lensed counts above a flux threshold $S$ is
$n(>S)=\int_{S}^{+\infty} dS^{\prime} \mu^{-2} f(S^{\prime}\mu^{-1})$
(see Eq.~[\ref{eq:dnds_mu}]). After expanding in powers of $\epsilon$
and integrating by parts, we obtain, to first order in $\epsilon$,
\begin{equation}
n(>S) \approx \hat{n}(>S)- \epsilon 
\left[ \hat{n}(>S) - \left. \frac{d\hat{i}}{d\hat{S}} \right|_{S} \right],
\label{eq:n_epsilon}
\end{equation}
where we have used the unlensed version of equation~(\ref{eq:dids}).  The
tendency of $n(>S)$ to increase or decrease with $\mu$ therefore
depends on the sign of the quantity in brackets. This sign can be
conveniently determined by comparing the solid line in figure
\ref{fig:nis_s}b with that in figure \ref{fig:nis_s}c.  A cross-over
between the solid and the weak lensing equivalent of the dashed and
dot-dashed lines in figure \ref{fig:nis_s}b will occur at any values
of $S=S_{*}$ for which $\hat{n}(>S_{*})=\left. ({d\hat{i}}/{d\hat{S}})
\right|_{S_{*}}$. This result holds for any function $f(\hat{S})$, as
long as $\hat{n}(>S)$ converges.

\subsection{Unresolved Intensity}
We now turn to the unresolved intensity which, in our notation, is
$i(<S)=\int_{0}^{S}dS^{\prime} S^{\prime} \mu^{-2}
f(S^{\prime}\mu^{-1})$ (see Eq.~[\ref{eq:dids}]).  As before, an
expansion in powers of $\epsilon$ and an integration by parts yields
to leading order in $\epsilon$,
\begin{equation}
i(<S) \approx \hat{i}(<S)- \epsilon 
\left[ \left. \frac{d \hat{\sigma}_{i}^{2}}{d \hat{S}} \right|_{S} \right],
\label{eq:i_epsilon}
\end{equation}
where $\left. ({d \hat{\sigma}_{i}^{2}}/{d \hat{S}}) \right|_{S} \equiv
S^2 f(S)$ (see Eq.~[\ref{eq:dsigi2ds}]).  
In analogy with
equation~(\ref{eq:n_epsilon}), the quantity in bracket in
equation~(\ref{eq:i_epsilon}) involves a higher moment, namely
$\left. ({d \hat{\sigma}_{i}^{2}}/{d \hat{S}}) \right|_{S}$, which can be
determined from the solid line in
figure~\ref{fig:nis_s}e. However, the quantity in brackets in
equation~(\ref{eq:i_epsilon}) does not contain a term analogous to the first
term in the bracket of equation~(\ref{eq:n_epsilon}). This is due to a
cancellation which is unique to $i$. In general, we may consider any
integrated quantity $g$ of the form $g(S_{1},S_{2}) \equiv
\int_{S_{1}}^{S_{2}} dS S^{\lambda} \left. ({dn}/{dS})\right|_{S}$,
where $\lambda$ is a constant. It then follows that the cancellation
will occur only for $\lambda=1$, i.e. for $g \equiv i$. This
feature of $i$ is also responsible for the conservation of the
total integrated intensity $i(0,\infty)$ under magnification, which was
discussed in \S\ref{effect_lensing_general}.  As a result, the
quantity in square brackets in equation~(\ref{eq:i_epsilon}) is 
positive definite, and so $i(<S)$ {\it always} decreases as $\mu$ increases.
Magnification
always causes sources to appear brighter and thus allows
a larger fraction of the background to be resolved.
Consequently, no cross-over occurs between the solid and
dot-dashed or dashed lines in figure~\ref{fig:nis_s}d.

A similar analysis shows that 
the above cancellation does not occur
for $\sigma_{i}^{2}(<S)$ (corresponding
to $\lambda=2$). Cross-overs are
allowed in this case, as shown in figure~\ref{fig:nis_s}f.
Again, these results are valid for any function $f(\hat{S})$ as long
as the integrated intensity and variance converge.

\section{Count Statistics}
\label{app:stat}
\subsection{Source Counts}
\label{app:stat_n}
In this appendix, we evaluate the statistical significance of the
lensed source counts as compared to the unlensed counts. Let us
consider source counts in $n_{R}$ concentric rings $R_{j},
j=1,\dots,n_{R}$ centered on the cluster center. We denote the mean
number of sources in $R_{j}$ as $\overline{N}_{j}$ and
$\hat{\overline{N}}_{j}$ in the lensed and unlensed case,
respectively.

In a single experiment, one measures the number of sources $N_{j}$ in each
ring $R_{j}$. The variable $N_{j}$ follows a Poisson distribution with mean
$\overline{N}_{j}$ and $\hat{\overline{N}}_{j}$, in the lensed and unlensed
case. The mean unlensed counts $\{\hat{\overline{N}}_{j}\}$ can be
determined from a large-area observation of the XRB ``in the field'' (i.e.
away from any cluster) with an accuracy which is effectively arbitrary, and
can thus be treated as well determined constants. We want to test and
possibly rule out the hypothesis that the measured counts $\{N_{j}\}$ were
drawn from the unlensed distribution. For this purpose, we treat each ring
as an independent measurement and consider the following statistic in a
single experiment
\begin{equation}
X^{2}_{\rm single}(N_{k};\hat{\overline{N}}_{k}) \equiv
\sum_{j=1}^{n_{R}} \frac{(N_{j}-\hat{\overline{N}}_{j})^2}
{\hat{\overline{N}}_{j}}.
\end{equation}
Since the considered distributions are Poisson and not Gaussian, it
is not guaranteed that $X^{2}_{\rm single}$ follows a $\chi^{2}$
distribution. Numerical simulations reveal, however, that, for our
experimental conditions, $X^{2}_{\rm single}$ does follow this
distribution to a good approximation.  This is true even when $n_{R}$
is as low as 2 and when $\hat{\overline{N}}_{k}$ is as low as 4. In our
case, we can thus derive likelihood probabilities using the usual
$\chi^2$ tables.

Our probability estimate is obtained by averaging $X^{2}_{\rm single}$
over a large number $n_{E}$ of experiments. If we denote
by $N_{k,e}$ the source count in $R_{k}$ for the e$^{\rm th}$ 
experiment, the mean statistic is
\begin{equation}
X^{2}(\overline{N}_{l};\hat{\overline{N}}_{l}) \equiv
\frac{1}{n_{E}} \sum_{e=1}^{n_{E}}
X^2_{\rm single}(N_{k,e},\hat{\overline{N}}_{k}),
\end{equation}
which can be easily evaluated numerically.

As an example, let us consider the third, fourth and fifth rings in
figure~\ref{fig:n_rings} which we rename as $R_{1}$, $R_{2}$, and
$R_{3}$, respectively. In this case, $\overline{N}_{1} \approx 19.3$,
$\overline{N}_{2} \approx 404.4$, and $\overline{N}_{3} \approx
4426.7$, whereas $\hat{\overline{N}}_{1} \approx 28.7$,
$\hat{\overline{N}}_{2} \approx 444.8$, and $\hat{\overline{N}}_{3}
\approx 4561.7$. Note that although $\mu \propto \theta^{-1}$ at
large angles, the values of $\hat{\overline{N}}_i$ do not follow a
geometric sequence because of the variable detection threshold
(see \S\ref{variable_threshold}). 
By numerically averaging over $n_{E}=10^3$ experiments, we obtain $X^2
\approx 13.38$. The $\chi^2$-probability to exceed this value is about
99.61\% for 3 degrees of freedom. The lensed counts determined in a
single experiment are thus, on average, distinguishable from the
unlensed counts at the $2.9\sigma$ significance level.

\subsection{Photon Counts}
\label{app:stat_p}
In this section, we quantify the significance of the reduction in the
photon counts due to lensing after the resolved sources have been removed.
For this purpose, let $\overline{P}_{j}$ be the mean of the total number of
photons (i.e. the sum of the cluster emission and the lensed unresolved
XRB) in a ring $R_{j}$, and let $\sigma_{P,j}$ be the associated standard
deviation.  The number of photons expected in the same ring for the
unlensed unresolved XRB is denoted by $\hat{\overline{P}}_{j}$.

Again, we first consider a single experiment in which $P_{j}$ photons
are detected in the ring $R_{j}$. Here, a useful statistic to consider is
\begin{equation}
X^2_{\rm single}(P_{k};\sigma_{P,k};\hat{\overline{P}}_{k}) \equiv
\sum_{j=1}^{n_R} \frac{(P_{j}-\hat{\overline{P}}_{j})^2}
                      {\sigma_{P,j}^{2}}.
\end{equation}
In general, the distributions of the variables $P_{j}$ are not Gaussian. 
We have indeed seen in \S\ref{obs_unresolved} that their distribution
results from a combination of intrinsic background fluctuations and photon
shot noise for both the cluster and XRB contributions. In our
case, however, the mean photon counts are large (typically
$\overline{P}_{k} \gtrsim 2\times10^4$ for the outer rings)
and the distribution is close to Gaussian. We can therefore take
$X^2_{\rm single}$ to be distributed as a $\chi^2$-variable
with $n_{R}$ degrees of freedom.

\begin{figure}
\plotone{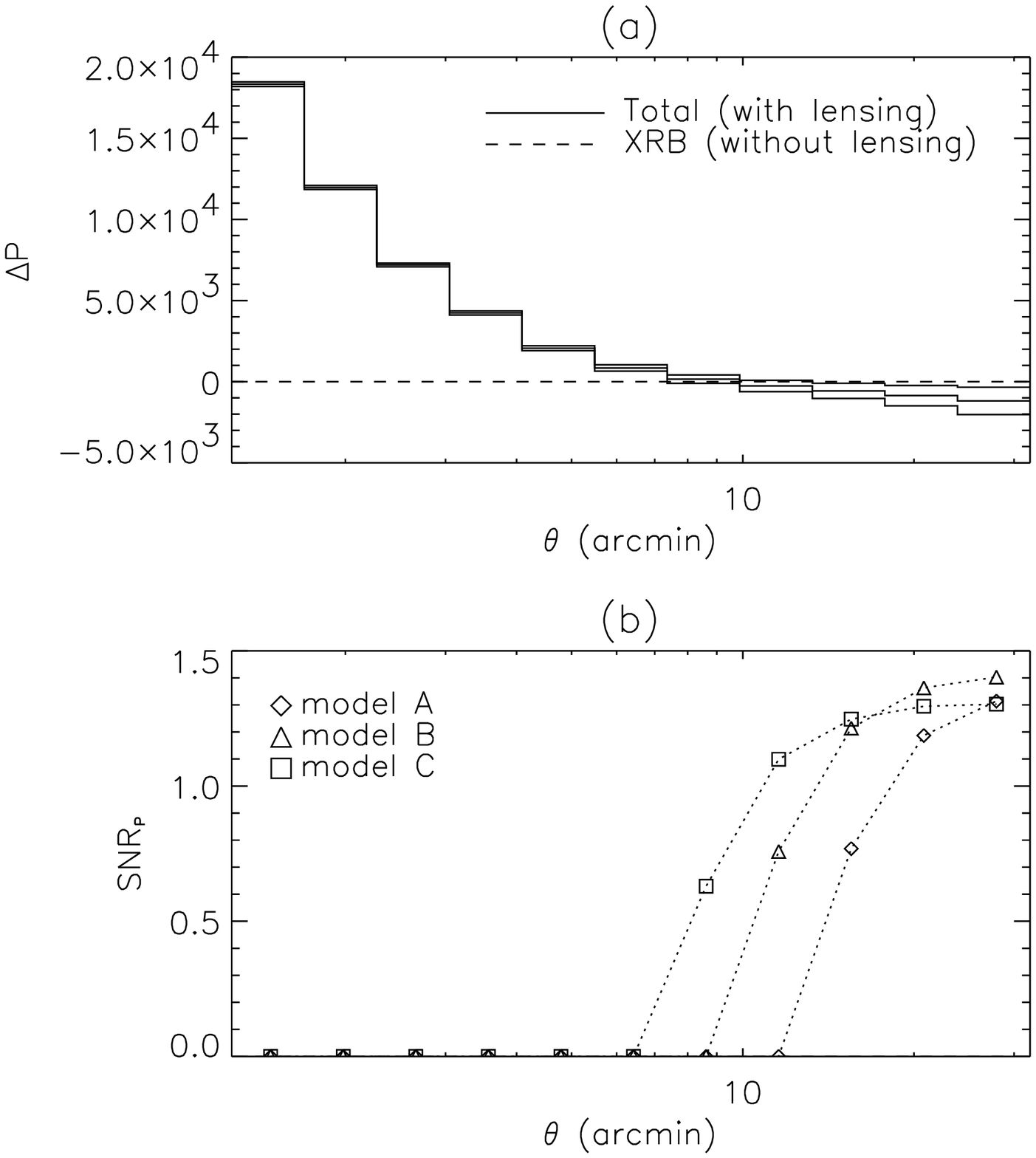}
\caption{Effect on lensing on the unresolved component of the XRB for
a $10^6$ s observation of A1689 with the AXAF-ACIS camera. For clarity, the
angular range was extended to $\theta=32'$, i.e. to four times the field of
view of this instrument. The flux detection threshold was set to
$S_r(0.5-2\mbox{keV})=1, 6, 30 \times 10^{-17}$ ergs cm$^{2}$ s$^{-1}$ for
model A, B, and C, respectively, and $b$ was set to a value of 0.8 . Panel
(a) shows the photon count difference $\Delta P$ for model B. The central
solid line shows the mean count difference, whereas the two neighboring
lines correspond to a single standard deviation $\sigma_{P,tot}$ away from
the mean. The dashed line corresponds to $\Delta P = 0$, i.e. to an
observation of the (unlensed) XRB ``in the field''. Panel (b) shows the
signal-to-noise ratio SNR$_{P}$ for each ring and for each XRB model.
\label{fig:i_rings}}
\end{figure}

If we perform a large number $n_{E}$ of experiments and denote by
$P_{k,e}$ the photon counts in $R_{k}$ for the e$^{\rm th}$ experiment,
then the mean statistic is
\begin{equation}
X^2(\overline{P}_{l};\sigma_{P,l};\hat{\overline{P}}_{l}) 
\equiv \frac{1}{n_{E}} \sum_{e=1}^{n_{E}} 
X^2_{\rm single}(P_{k,e};\sigma_{P,k};\hat{\overline{P}}_{k}).
\end{equation}
As an example, let us rename the ninth, tenth, eleventh ring in
figure~\ref{fig:i_rings} as $R_{1}$, $R_{2}$, and $R_{3}$,
respectively. In this case, $\overline{P}_{1} \approx 34621$,
$\overline{P}_{2} \approx 62696$, $\overline{P}_{3} \approx 112693$,
$\hat{\overline{P}}_{1} \approx 35189$, $\hat{\overline{P}}_{2}
\approx 63556$, $\hat{\overline{P}}_{3} \approx 113879$, $\sigma_{P,1}
\approx 468$, $\sigma_{P,2} \approx 631$, $\sigma_{P.3} \approx
846$. This results in $X^2 \approx 8.43$, which corresponds to a
$\chi^2$ probability of 96.2\% for 3 degrees of freedom. The total
lensed photon counts in a single experiment are thus, on average,
below the unlensed XRB with a significance of $2.0\sigma$.



\clearpage


\begin{deluxetable}{rrrrr}
\tablecaption{Parameters for models of the logN-logS relation
for X-ray sources. \label{tab:xrbmodels}}
\tablewidth{0pt}
\tablehead{
\colhead{} &
\colhead{Model A} &
\colhead{Model B} &
\colhead{Model C} &
\colhead{ROSAT\tablenotemark{a}}
}
\startdata
$\eta_1\tablenotemark{b}$   &  1.98e-22  &  1.98e-22  &  1.98e-22  & 1.98e-22 \nl   
$\eta_2\tablenotemark{b}$   &  7.45e-09  &  9.36e-11  &  4.11e-12  & 7.68e-12 \nl
$\eta_3\tablenotemark{b}$   &  7.45e-09  &  1.13e-06  &  9.19e+01  &   --    \nl
$\beta_1$                   &  2.72      &  2.72      &  2.72      & 2.72     \nl
$\beta_2$                   &  1.72      &  1.86      &  1.96      & 1.94     \nl
$\beta_3$                   &  1.72      &  1.60      &  1.10      &   --     \nl
$S_{12}\tablenotemark{c}$   &  2.66e-14  &  2.66e-14  &  2.66e-14  & 2.66e-14 \nl
$S_{23}\tablenotemark{c}$   &  2.00e-16  &  2.00e-16  &  2.00e-16  &
                            2.50e-15\tablenotemark{d} \nl
\%XRB\tablenotemark{e}      &  77\%      &  100\%     &  100\%     &   --     \nl
\enddata
\tablenotetext{a}{observed ROSAT counts (Hasinger et al. 1993)}
\tablenotetext{b}{Normalization in deg$^{-2}$ (ergs cm$^{-2}$ s$^{-1}$)$^{\beta-1}$}
\tablenotetext{c}{fluxes in ergs cm$^{-2}$ s$^{-1}$ in the 0.5-2 keV range} 
\tablenotetext{d}{Survey detection threshold}
\tablenotetext{e}{Fraction of the XRB from point sources for each model}
\end{deluxetable}



\begin{figure}
\figurenum{6b}
\plotone{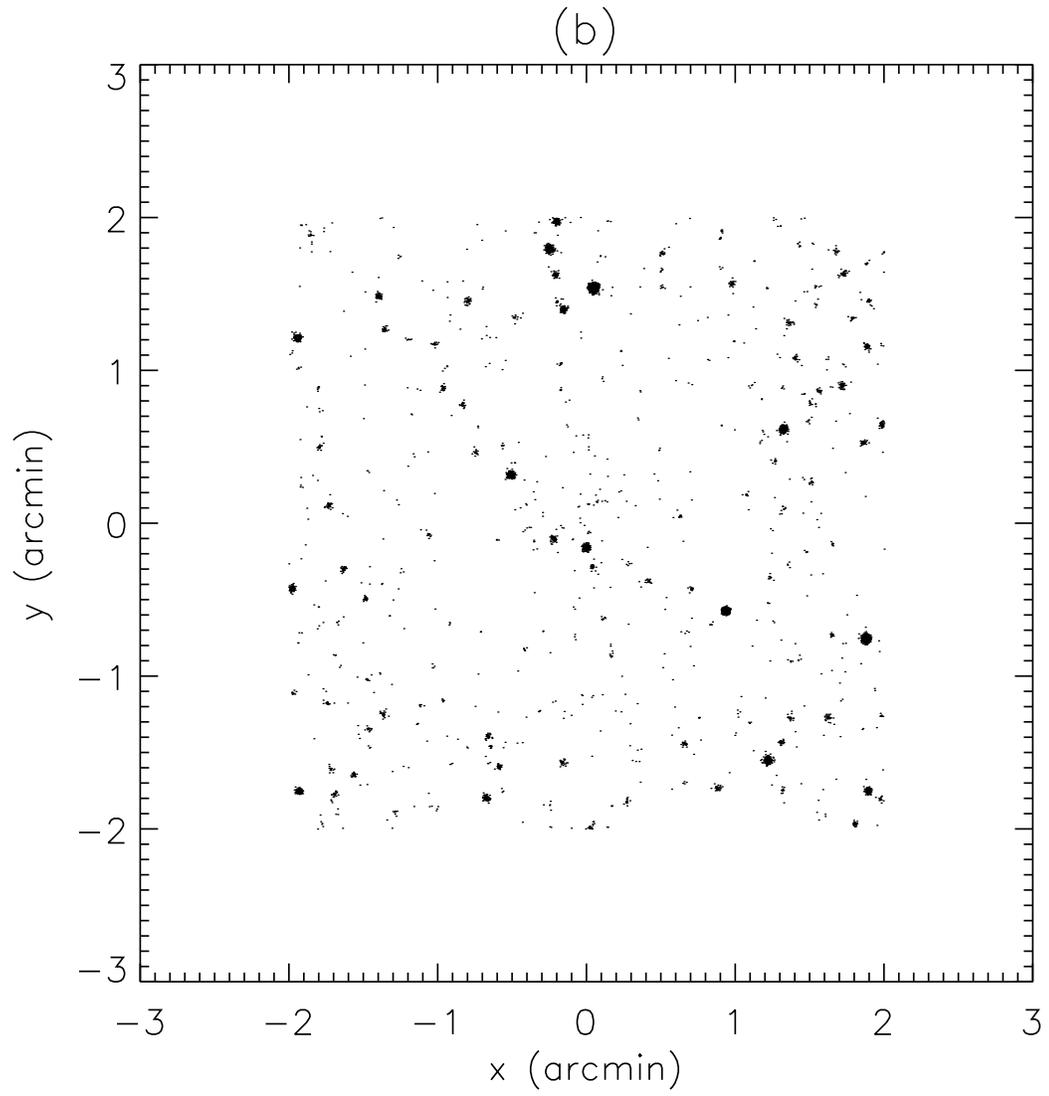}
\caption{[See caption on previous page]}
\end{figure}

\end{document}